\documentclass[longauth]{aa} 

\usepackage{graphicx}
\usepackage{txfonts}
\newcommand{\coco}{\texttt{CoCo}}

\newcommand{\magarc}{mag arcsec\ensuremath{^{\mathrm{-2}}}}
\newcommand{\mulim}{\ensuremath{\mu_{\rm lim}}}

\DeclareUnicodeCharacter{223C}{~}
\newcommand{\degree}{^\circ}

\newcommand{\detectionaperradarcsec}{7}

\newcommand{\detectionapernumrandom}{10000}
\newcommand{\detectionnckernelfwhm}{3}
\newcommand{\detectionnckerneltrunc}{4}
\newcommand{\detectionnckernelfwhmseg}{5}
\newcommand{\detectionnctilesize}{40}
\newcommand{\detectionncholesize}{10000}
\newcommand{\detectiongnuastrover}{0.14}

\newcommand{\detectionapersb}{26.81}
\newcommand{\detectionaperupsigma}{4.32}

\begin{document}

   \title{Hidden depths in the local Universe: The Stellar Stream Legacy Survey}


\author{David Mart{\'\i}nez-Delgado \inst{1,2}\thanks{Talentia Senior Fellow}, Andrew\ P.\ Cooper\inst{3,4,27}\thanks{Yushan Fellow},
 Javier Rom{\'a}n\inst{1,11,12}, Annalisa Pillepich\inst{5}, Denis Erkal\inst{6}, Sarah Pearson\inst{7}\thanks{Hubble Fellow}, John Moustakas\inst{8}, Chervin F. P. Laporte\inst{9}, Seppo Laine\inst{10}, Mohammad Akhlaghi\inst{11,12},  Dustin Lang\inst{13}, Dmitry Makarov\inst{14}, Alejandro S. Borlaff\inst{15},  Giuseppe Donatiello\inst{16}, William\ J.\ Pearson\inst{17}, Juan Mir\'o-Carretero\inst{18},  Jean-Charles Cuillandre\inst{19}, Helena Dom\'\i nguez\inst{20}, Santi Roca-F\`abrega\inst{18}, Carlos S. Frenk\inst{21}, Judy Schmidt\inst{22},  Mar\'\i a A. G\'omez-Flechoso\inst{18}, Rafael Guzman\inst{18}, Noam I. Libeskind\inst{23}, Arjun Dey\inst{24}, Benjamin A. Weaver\inst{24}, David Schlegel\inst{25}, Adam D. Myers\inst{26}, Frank G. Valdes\inst{24}}

   \institute{Instituto de Astrof\'isica de Andaluc\'ia, CSIC, Glorieta de la Astronom\'\i a, E-18080, Granada, Spain 
   \and
   Astronomisches Rechen-Institut, Zentrum f\"ur Astronomie der
Universit\"at Heidelberg, M\"onchhofstr.\ 12--14, 69120 Heidelberg,
Germany
\and
Institute of Astronomy and Department of Physics, National Tsing Hua University, Kuang Fu Rd. Sec. 2, Hsinchu 30013, Taiwan
\and
Center for Informatics and Computation in Astronomy, National Tsing Hua University, Kuang Fu Rd. Sec. 2, Hsinchu 30013, Taiwan
\and
Max Planck Institut f\"{u}r Astronomie, Heidelberg, Germany
\and
Department of Physics, University of Surrey, Guildford GU2, 7XH, UK 
\and
Center for Cosmology and Particle Physics, Department of Physics, New York University, 726 Broadway, New York, NY 10003, USA
\and
Siena College, Department of Physics \& Astronomy, 515 Loudon Road, Loudonville, NY, 12211, USA
\and
Kavli Institute for the Physics and Mathematics of the Universe (WPI), The University of Tokyo, Chiba 277-8583, Japan
\and
IPAC, Mail Code 314-6, Caltech, 1200 E. California BLVd., Pasadena, CA 91125 USA
\and
Instituto de Astrof\'isica de Canarias (IAC), Calle V\'ia L\'actea s/n, E-38205 La Laguna, Tenerife; Spain 
\and
Facultad de F\'isica, Universidad de La Laguna, Avda. Astrof\'isico Fco. S\'anchez s/n, 38200La Laguna, Tenerife, Spain. 
\and
Perimeter Institute for Theoretical Physics,
31 Caroline St N, Waterloo, Canada
\and
Special Astrophysical Observatory of the Russian Academy of Sciences, Nizhnij Arkhyz, 369167, Russia
\and
NASA Ames Research Center, Moffett Field, CA 94035, USA
\and
UAI - Unione Astrofili Italiani /P.I. Sezione Nazionale di Ricerca Profondo Cielo, 72024 Oria, Italy 
\and
National Centre for Nuclear Research, ul. Pasteura 7, 02-093 Warszawa, Poland 
\and
Departamento de F{\'\i}sica de la Tierra y Astrof{\'\i}sica, Universidad Complutense de Madrid, E-28040 Madrid, Spain
\and
AIM, CEA, CNRS, Université Paris-Saclay, Université de Paris, F-91191 Gif-sur-Yvette, France
\and
Institute of Space Sciences (ICE,CSIC), Campus UAB, Carrer de Magrans, E-08193 Barcelona, Spain
\and
Institute for Computational Cosmology, University of Durham, UK 
\and
Astrophysics Source Code Library, University of Maryland, 4254 Stadium Drive College Park, MD 20742, USA 
\and
Leibniz-Institut f\"{u}r Astrophysik Postdam (AIP), An der Sternwarte 16, D-14482 Postdam, Germany
\and
NSF's National Optical-Infrared Astronomy Research Laboratory, 950 N. Cherry Ave., Tucson, AZ, 85719
\and
Lawrence Berkeley National Laboratory, 1 Cyclotron Rd., Berkeley, CA 94720
\and
Department of Physics \& Astronomy, University of Wyoming, 1000 E. University, Dept. 3905, Laramie, WY 82071, USA
\and
Physics Division, National Center for Theoretical Sciences, Taipei 10617, Taiwan
}

   \date{}

\authorrunning{Mart{\'\i}nez-Delgado et al.}
\titlerunning{The Stellar Stream Legacy Survey}
 
  \abstract
   {Mergers and tidal interactions between massive galaxies and their dwarf satellites are a fundamental prediction of the Lambda-cold dark matter cosmology. These events are thought to provide important observational diagnostics of non-linear structure formation. Stellar streams in the Milky Way and Andromeda are spectacular evidence for ongoing satellite disruption. However, constructing a statistically meaningful sample of tidal streams beyond the Local Group has proven a daunting observational challenge, and the full potential for deepening our understanding of galaxy assembly using stellar streams has yet to be realised.}
   {Here we introduce the {\it Stellar Stream Legacy Survey}, a systematic imaging survey of tidal features associated with dwarf galaxy accretion around a sample of $\sim3100$ nearby galaxies within $z\sim0.02$, including about $940$ Milky Way analogues.}
   {Our survey exploits public deep imaging data from the {\it DESI Legacy Imaging Surveys}, which reach surface brightness as faint as $\sim$ 29 \magarc\ in the $r$ band. As a proof of concept of our survey, we report the detection and broad-band photometry of 24 new stellar streams in the local Universe.}
   {We discuss how these observations can yield new constraints on galaxy formation theory through comparison to mock observations from cosmological galaxy simulations. These tests will probe the present-day mass assembly rate of galaxies, the stellar populations and orbits of satellites, the growth of stellar halos, and the resilience of stellar disks to satellite bombardment.}
   {}


   \maketitle
%

\section{Introduction}
\label{sec:intro}

Within the hierarchical framework for galaxy formation, the stellar halos of massive galaxies are expected to form predominantly through a succession of mergers with lower-mass satellite galaxies. Cosmological numerical simulations of galaxy assembly in the  Lambda-cold dark matter ($\Lambda$CDM) paradigm predict that satellite disruption occurs throughout the lifetime of all massive galaxies \citep[e.g.][]{bullock_johnston_2005,delucia_helmi_2008,cooper_2010,cooper_2013,pillepich_2015,RodriguezGomez16}. As a consequence,  the stellar halos of massive galaxies at the current epoch should contain a wide variety of diffuse remnants of tidally disrupted dwarf satellites. 
The most spectacular examples of such debris are long, dynamically cold stellar streams\footnote{Features arising from tidal disruption are often classified by their visual morphology, for example tidal tails or arms from mergers of the host galaxy with a gas rich, disk-like companion \citep{byrd_howard_1992}; fans and plumes from `dry' major mergers \citep{feldmann_2008} and shells around massive early-type galaxies \citep{schweizer_seitzer_1992,cooper_2011}. Here we follow the terminology of \citet{duc_2015}.} that wrap around the host galaxy and roughly trace the orbit of the progenitor satellite \citep[e.g.][]{sanders_binney_2013}. Although these fossils of galactic cannibalism disperse into amorphous clouds of stars (through phase-mixing) in a few gigayears, $\Lambda$CDM simulations predict that significant numbers of coherent tidal features may be detectable with sufficiently deep observations in the outskirts of the majority of nearby massive galaxies \citep{font_2011,cooper_2013,pillepich_2015}.

The detection and characterisation of the faint remnants of dwarf galaxy accretion is a vital test of the hierarchical nature of galaxy formation \citep[][]{johnston_etal_2008} which has not yet been fully exploited, mainly because their extremely low surface brightness makes them challenging to observe. Although some of the most luminous examples of tidal tails and shells around massive elliptical galaxies have been known for many decades \citep[e.g.][]{arp_1966,malin_carter_1980}, recent studies have shown that fainter analogues of these structures are common around spiral galaxies in the local Universe, including the Milky Way and Andromeda. The first known galactic tidal stream around the Milky Way (the Sagittarius stream) was discovered almost three decades ago \citep{ibata94,mateo_1998,martinezdelgado_2001,majewski_2003}. In recent years, 
a new generation of wide-field digital imaging surveys (including the Sloan Digital Sky Survey, Pan-STARRS, and the Dark Energy Survey) 
have revealed a multitude of fainter streams \citep{belokurov_2006,shipp_2018}
and other amorphous sub-structures in our Galaxy's stellar halo \citep[e.g.][]{newberg02,juric_2008,bernard_2016}.
The Pan-Andromeda Archaeological Survey \citep[PAndAS;][]{ibata_2007,mcconnachie_2009} has provided an equally revolutionary panoramic view of 
similar features in the stellar halo of M31.
These observations provide sound empirical support for the $\Lambda$CDM prediction that dwarf galaxy disruption is responsible for the streams and diffuse stellar halos around massive late-type galaxies.  

A search for comparable fossil records 
in a 
much larger sample of nearby galaxies is required to allow for the possibility that the recent merging histories of the two Local Group spirals may not be `typical' \citep[e.g.][]{mutch_2011}. Improvements in $\Lambda$CDM numerical simulations have helped to guide the quest 
to discover and characterise stellar streams
\citep[e.g.][]{bullock_johnston_2005,cooper_2010}. Recent simulations have demonstrated that the characteristics of sub-structures similar to those 
observed are sensitive to the recent merger histories of their host galaxies.
Models predict that a survey of $\sim$100 parent galaxies reaching a surface brightness limit of $\sim$29 \magarc\ would reveal many tens of tidal features, with approximately one detectable feature per galaxy \citep[][]{johnston_2001}. 
One of the core motivations for the survey we describe here is that the statistical comparison of these detailed simulations with observations has not yet been possible, because no suitably deep data are available for a representative sample of galaxies.

The primary reason for the highly incomplete observational portrait of dwarf satellite accretion events is the inherent difficulty of detecting the low surface brightness features they create, even in the local Universe. Only a few instances of extragalactic stellar tidal streams had been reported around other nearby galaxies before the beginning of this century. Using contrast enhancement techniques on deep photographic plates, \citet{malin_hardley_1997} highlighted two possible tidal streams surrounding the galaxies M83 and M104. 
In the last decade, the way forward has been demonstrated by a programme of deep, 
 wide-field images of nearby Milky-Way analogue galaxies in the local volume taken with small telescopes \citep[{the \it Stellar Tidal Stream Survey}, STSS;][]{martinezdelgado_2019}. This survey has revealed, in many cases for the first time, an assortment of large-scale tidal structures around nearby spiral galaxies, with striking morphological characteristics that are qualitatively consistent with the predictions of $N$-body simulations \citep[][see their Fig.\ 2]{johnston_etal_2008,martinezdelgado_2010}. Further deep imaging surveys of the outskirts of local galaxies have been completed over the past decade \citep{tal_2009,ludwig_2012,duc_2015,laine_2014,rich_2019} or are currently ongoing \citep{kado_2018}. Resolved stellar population studies of stellar halo substructure \citep[similar to PAndAS,][]{mcconnachie_2009, mcconnachie2018} are only feasible for a select number of nearby galaxies within a distance of a few Megaparsecs using very large telescopes \citep[][]{mouhcine_ibata_2009,crnojevic_2017} or from space \citep[][]{mouhcine_2010,crnojevic_2016,carlin_2019}.

Many of the studies to date have targeted spiral galaxies with known features visible in shallower amateur imaging or the Sloan Digital Sky Survey \citep[SDSS,][]{abazajian_2009}. However, subjective target selection criteria,
such as the morphological type of the host galaxy, make
it challenging to compare the extremely heterogeneous samples currently available with one another, or with predictions from
cosmological simulations. Therefore, a meaningful comparison between data and models demands deep imaging data for a large number of galaxies with a simple, quantitative selection function, based on fundamental quantities such as the stellar mass of the host galaxy. 
Although several studies listed above have taken important steps towards this goal, there are significant differences between their results, likely due to different sample selections, detection criteria and surface brightness limits \citep[see][for a recent discussion]{hood_2018}.

In this paper we present the new {\it Stellar Stream Legacy Survey}. This survey focuses on stellar tidal features formed during the accretion of low-mass dwarf galaxies orbiting host galaxies of luminosity comparable to the Milky Way. For brevity, we refer to all such features as \textit{stellar tidal streams}, or simply \textit{streams}, regardless of their detailed morphology. This definition excludes tidal tails and other structures arising from mergers with a characteristic mass ratio greater than 1:10, in which the gravitational potential of the host is strongly perturbed. In contrast to the study of major merger remnants, the systematic study of `mini-merger' events (with a characteristic mass ratios of $\sim$ 1:50 -- 1:100) is a relatively new area of exploration. 

Due to their extremely faint surface brightness, clear examples of stellar streams are rarely detected even in ultra-deep imaging surveys. In a recent systematic search combining SDSS DR9 images in multiple bands, \citet{morales_2018} detected streams around only $\sim 10\%$ of a mass-selected sample of isolated Milky Way analogues (see Sec.\ 2.2). Moreover, the majority of previously known streams were only barely visible in the deep images constructed by \citet{morales_2018}, making it infeasible to analyse their photometric and structural properties from SDSS data alone. The exploration of a few thousand galaxies with deeper data is needed to obtain the uniform, statistically representative samples required for meaningful comparisons to theory. Undertaking this type of survey with small robotic telescopes would be very expensive, given the substantial fraction of non-detections even in the ultra-deep imaging ($\sim$ 29.0 \magarc) obtained with such facilities (see Sec.~\ref{sec:othersurveys}).

The promising results of previous forays into this topic argue strongly that a comprehensive study of tidal streams in the nearby universe provide a new perspective on galaxy formation. With the {\it Stellar Stream Legacy Survey} we aim to collect a systematic deep, broad-band images of stellar streams around a much larger sample of galaxies than any previous survey of such features. This is now possible thanks to the latest generation of large-area imaging surveys. The quality of these data offers a unique opportunity to reveal as-yet hidden details of galaxy assembly in the local Universe and to make meaningful connections with theoretical predictions from the $\Lambda$CDM paradigm. 

In Sec.~\ref{sec:survey} we describe the scientific motivation and design of our {\it Stellar Stream Legacy Survey}, starting with an overview of the underlying data from the {\it DESI Legacy Imaging Surveys} \citep[][]{dey_2019} in Sec.~\ref{sec:legacy}. In Sec.~\ref{sec:cutout_method} we describe how we extend the reduction pipeline of the Legacy Surveys to produce image cutouts and sky background subtraction suitable for the discovery and photometry of very low surface brightness features. In Sec.~\ref{sec:surveydesign} we explain how we select a parent sample of target galaxies in the Local Universe. To validate our approach to discovery and photometry, in Sec.~\ref{sec:analysis} we present a photometric analysis of 24 stream candidates spanning the range of distance, surface brightness and morphology in our parent sample. This section introduces our baseline approach to aperture measurements of surface brightness and colour. Even our small proof-of-concept sample adds significantly to the number of robust colour measurements for low-mass streams in the literature; we briefly discuss our findings in the context of the colour distribution of dwarf satellite galaxies. Sections~\ref{sec:confusion} and \ref{sec:cirrus} discuss sources of confusion associated with stream detections, with which any future statistical analysis will have to contend. Having estimated our surface brightness limit, we compare our survey with others in Sec.~\ref{sec:othersurveys}. In Sec.~\ref{sec:discussion} we look ahead to improvements in the theoretical techniques that are necessary to exploit our full sample. We pay particular attention to requirements for meaningful comparisons with cosmological simulations, and show two examples of how state-of-the-art simulations can be used to produce mock images tailored to our survey. We conclude in Sec.~\ref{sec:conclusions}.
All magnitudes are in the AB system \citep{oke_1983} unless otherwise noted.


\begin{figure*}
    \centering
 	\includegraphics[width=0.95\textwidth]{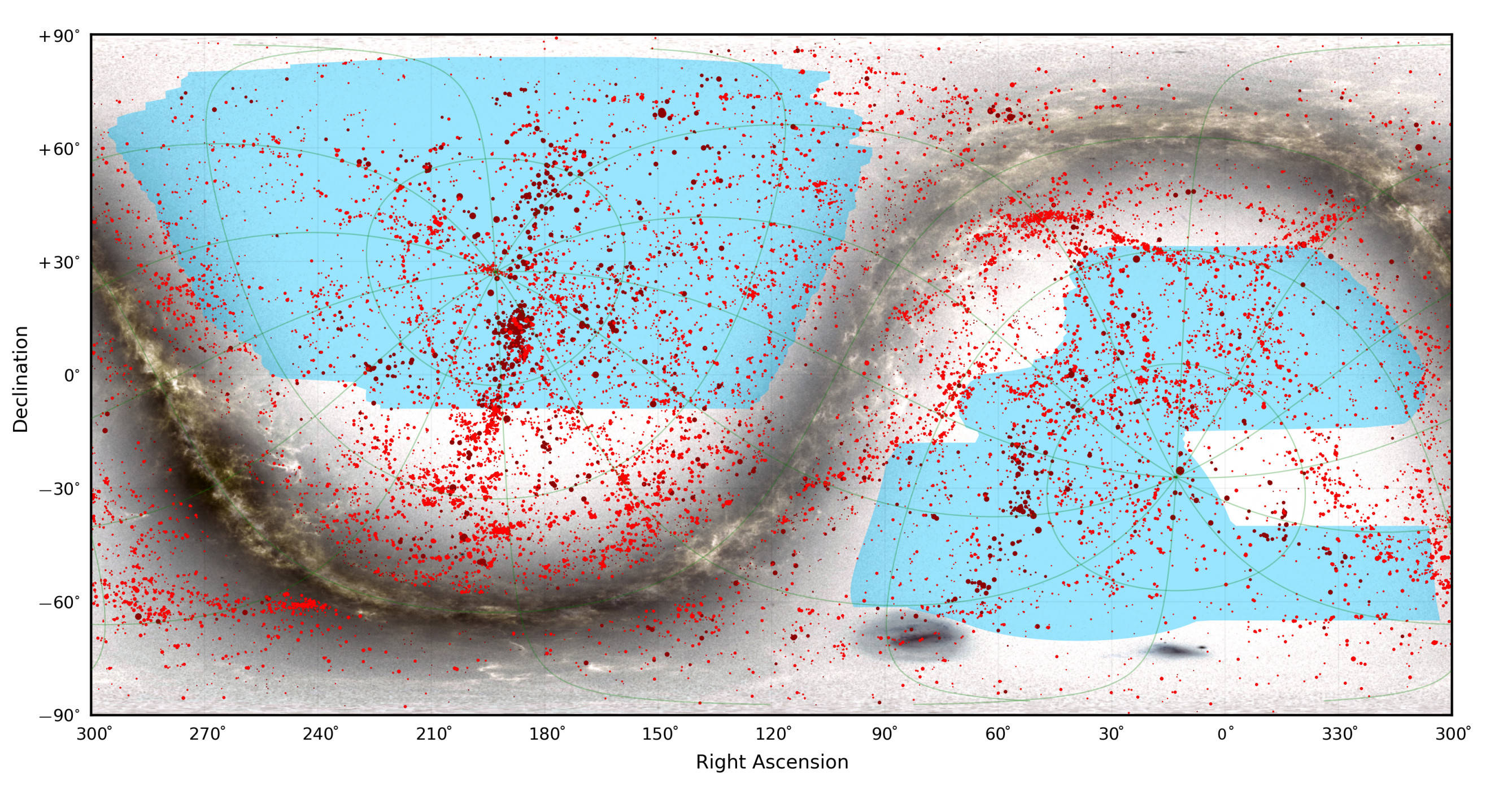}
	\includegraphics[width=0.95\textwidth]{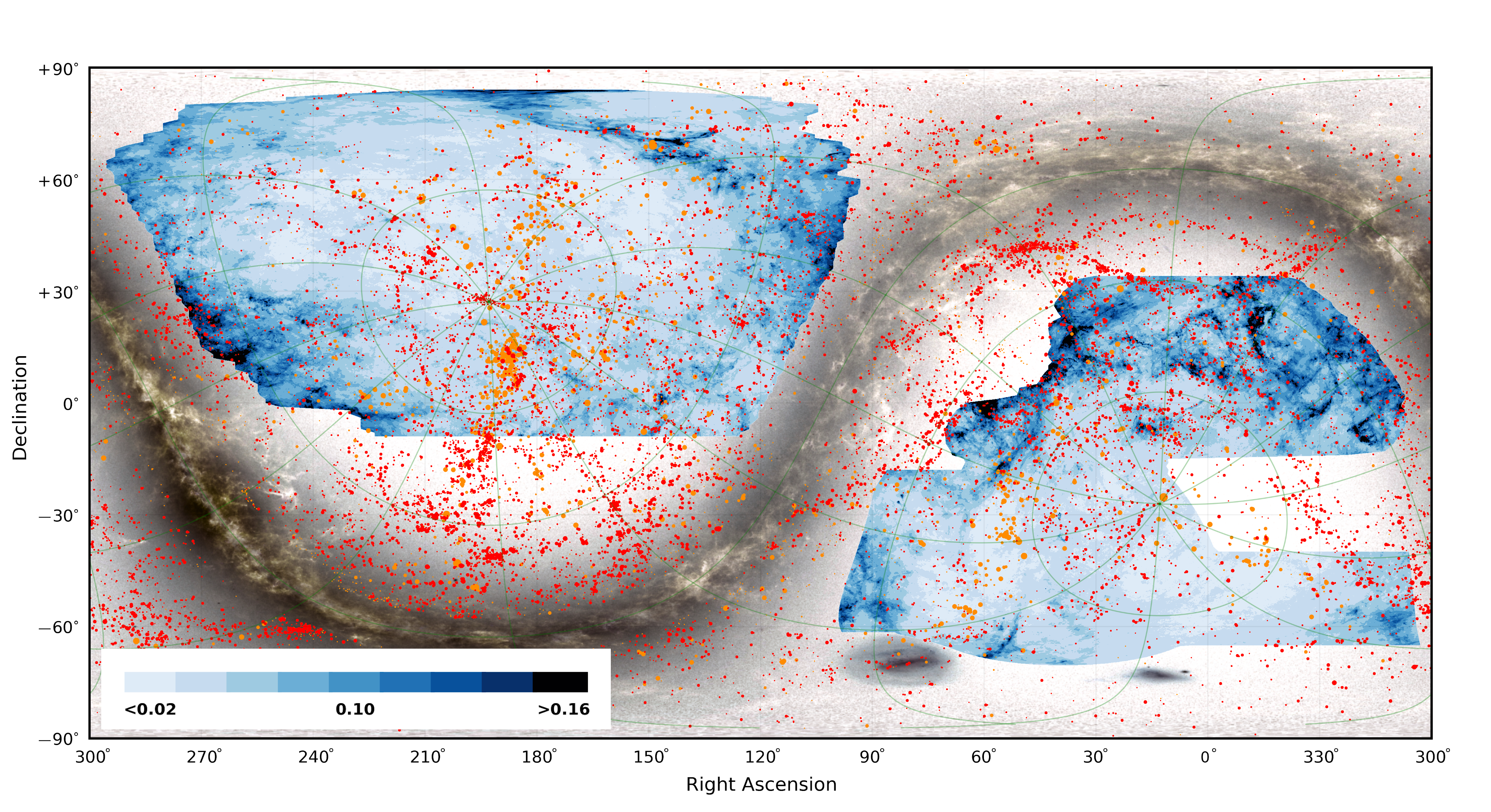}

           \caption{The {\it Stellar Stream Legacy Survey} sky coverage. ({\it Top}) The {\it DESI Legacy Imaging Surveys} footprint (blue), the source of optical data for our survey. The 2MASS Redshift Survey catalogue \citep{huchra_2012} beyond the local group and below $\mathrm{V}_{\rm LG} < 7000$ km s${^{-1}}$ is plotted over the whole sky with a marker size scaled to the $K$-band magnitude. Those galaxies matching our selection criteria ($2000< \mathrm{V}_{\rm LG} < 7000$ km s${^{-1}}$, $K$-band absolute magnitude M$_{K}< -19.6$) are displayed in bright red colour. The {\it Stellar Stream Legacy Survey} sub-samples comprise approximately 3000 of these galaxies in bright red within the blue area. Low galactic latitude regions of the survey affected by the higher stellar density of the Galaxy are perceived with this Gaia overlay. The galactic referential is plotted in green. ({\it Bottom}) E(B-V) distribution across the {\it DESI Legacy Imaging Surveys} footprint \citep{miville_2014}. Areas of greater dust density (dark blue) are also contaminated with the Galactic light diffused by the dust grains (galactic cirrus) causing a slightly higher background and occasional overlapping structures at scales similar to the faint streams motivating this study. The same 2MRS galaxies are represented on this second map with a different colour scheme.}
    \label{fig-depthmap}
\end{figure*}

\section{The Stellar Stream Legacy Survey}
\label{sec:survey}

\subsection{The DESI Legacy Imaging Surveys}
\label{sec:legacy}


The {\it Stellar Stream Legacy Survey} is based on a custom re-reduction of public data from the {\it DESI Legacy Imaging Surveys} project \citep{dey_2019}\footnote{See \url{https://legacysurvey.org}}, which combines images from several new wide area photometric surveys, by using a single analysis pipeline.
This provides high quality multi-colour imaging  with significantly greater depth and resolution than previous surveys such as SDSS or Pan-STARRS. The diffuse light feature detection limit of co-added images from the {\it DESI Legacy Imaging Surveys} is similar to that previously obtained with amateur telescopes \citep{martinezdelgado_2010,martinezdelgado_2015}, but additionally provides broadband photometry in standard filters\footnote{The deep images with the robotic telescopes from the Stellar Tidal Stream Survey \citep{martinezdelgado_2019} were obtained with a non-standard luminance filter under mediocre seeing conditions; see e.g.\ Fig.\ 1 in \citet{martinezdelgado_2015}} (see Sec.\ref{sec:othersurveys}) and sub-arcsec seeing, which allows us to trace fine details of streams at larger distances. The deep images shown in this paper are based on data release 8 (DR8) of the {\it DESI Legacy Imaging Surveys}, although future work will exploit the slightly greater coverage, depth and improved reduction of the more recent DR9.

Fig.~\ref{fig-depthmap} shows in blue the sky coverage of the DESI Legacy Imaging Surveys. The combined footprint of its constituent surveys covers approximately 20,000 deg$^2$ of the extra-galactic sky, approximately bounded by $-60{\degree} < \delta < +84{\degree}$ in celestial coordinates and $|b| > 18{\degree}$ in Galactic coordinates. 


The {\it DESI Legacy Imaging Surveys} comprise data in three optical bands ($g$, $r$ and $z$) coupled with all-sky infrared imaging from {\it Wide-field Infrared Survey Explorer} (WISE) \citep{wright2010,meisner2019}. The optical data were obtained by separate imaging projects on three different telescopes: The DECam Legacy Survey (DECaLS), the Beijing-Arizona Sky Survey (BASS) and the Mayall $z$-band Legacy Survey (MzLS) \citep{zou_2019,dey_2019}. The DECaLS and MzLS surveys, and the data reduction efforts, were undertaken to provide targets for the Dark Energy Spectroscopic Instrument (DESI) survey \citep{desi_2016a,desi_2016b}. The {\it DESI Legacy Imaging Surveys} data releases also include re-reduced public DECam data from other projects, including the Dark Energy Survey (DES)  \citep[][]{abbott_2018}. 

DECaLS provides complete $g$, $r$ and $z$-band DECam imaging with excellent seeing over 6200 deg$^2$ of the SDSS/BOSS/eBOSS footprint \citep[][]{smee_2013,dawson_2016}, covering both the North Galactic Cap region at $\delta \leq +32\degree$ and the South Galactic Cap region at $\delta \leq +34\degree$. Both DECaLS and DES data were obtained with the {\it Dark Energy Camera} (DECam) mounted on the Blanco 4-m telescope, located at the Cerro Torrolo Inter-American Observatory \citep{flaugher2015}.

In the Northern Hemisphere, the {\it DESI Legacy Imaging Surveys} comprise images from the BASS and MzLS surveys. These surveys have imaged the Dec.\ $\geq +32\degree$ footprint of the upcoming Dark Energy Spectroscopic Instrument (DESI) survey, as a complementary program to DECaLS. BASS provides $g$ and $r$ coverage using the 90Prime camera at the prime focus of the Bok 2.3m telescope located on Kitt Peak, adjacent to the 4-m Mayall Telescope. The 90Prime instrument is a prime focus 8k$\times$8k CCD imager, yielding a 1.12 deg field of view, with a scale of 0.$\arcsec$45 pixel$^{-1}$ \citep{williams2004}.  The survey has $g$- and $r$-band photometry over 5000 deg$^2$. The BASS survey tiled the sky in three passes, similar to the DECaLS survey strategy. At least one pass was observed in both photometric conditions and seeing conditions better than 1.7$\arcsec$.
The Mayall $z$-band Legacy Survey (MzLS) imaged the same sky area with a similar observing strategy, using the MOSAIC-3 camera \citep{dey_2016} at the prime focus of the 4m Mayall telescope at Kitt Peak National Observatory, in photometric conditions and with seeing better than 1.3 arcsec.

\subsection{Image cutouts and sky background subtraction}
\label{sec:cutout_method}

The unique low surface  brightness demands of our tidal stream survey requires bespoke sky-subtraction and co-addition of the calibrated survey images in the vicinity of each target galaxy. Here, we briefly describe our custom analysis, which utilises large portions of the \texttt{legacypipe}\footnote{Available at \texttt{https://github.com/legacysurvey/legacypipe}.} software infrastructure, the dedicated pipeline developed by the {\it DESI Legacy Imaging Surveys} team for source detection and model fitting of this survey imaging. One advantage of utilising \texttt{legacypipe} is that our custom analysis produces a standard and well-documented set of imaging and catalogue data products. Relative to the standard Legacy Survey co-added image products, these bespoke images reduce sky subtraction systematics around brighter host galaxies and so provide a more robust and uniform starting point for the visual detection of faint features, automated classification and quantitative photometry.

\begin{figure*}
	\includegraphics[width=\textwidth]{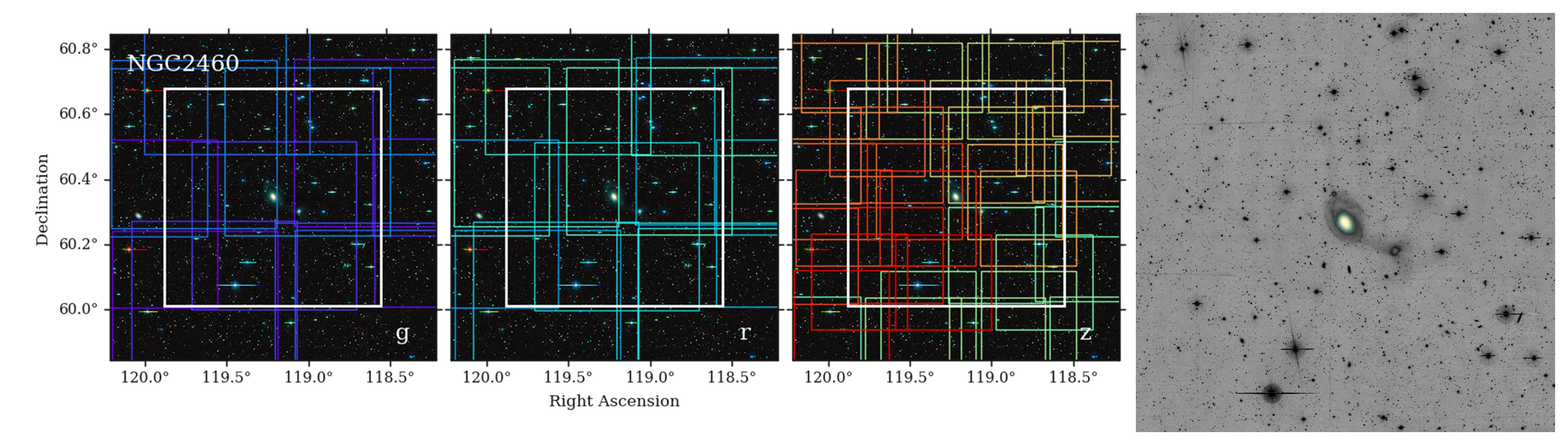}
          \caption{Example of the deep mosaics we construct from the individual {\it DESI Legacy Imaging Surveys} exposures overlapping our target galaxies. From left to right, panels show $g$, $r$, and $z$ bands. The white rectangle shows the region we process with our custom pipeline around a given target (here NGC 2460), coloured rectangles show each individual CCD frame in the stack. The overlapping tiling pattern is visible. Since this galaxy is in the northern footprint of the {\it DESI Legacy Imaging Surveys}, the $g$ and $r$ bands are from BASS, where the CCD pixel scale is larger than in the $z$ band, which is from MzLS. The right-most panel shows the final mosaic.}
  \label{fig-ccds}
\end{figure*}

We begin from the photometrically and astrometrically calibrated CCD images produced by the Legacy Surveys imaging team \citep[see][for details]{dey_2019}. Next, for the predefined custom footprint of each galaxy in our sample, we read the set of CCD exposures contributing to that field. Typically there are many separate partially overlapping CCD frames for a given target; an example is shown in Fig.~\ref{fig-ccds}. Then, we subtract the sky background from each CCD using a custom algorithm, described below, which preserves the low surface-brightness galactic features of interest. Finally, we use \textit{Lanczos3} (sinc) resampling to project each CCD onto a common (predefined) tangent plane and pixel scale ($0\farcs262$~pixel$^{-1}$), and combine all the data using inverse variance weights.

Our custom background subtraction algorithm consists of the following steps: We first bring all the input imaging in a given bandpass to a common (additive) pedestal by solving the linear equation $\mathbf{A}\mathbf{x}=\mathbf{b}$ for the full set of $N$ images and $M$ overlapping images, where $\mathbf{A}$ is an $M\times N$ matrix of the summed variances of overlapping pixels, $\mathbf{x}$ is an $N$-length vector of the additive offsets we must apply to each image (in calibrated flux units), and $\mathbf{b}$ is an $M$-length vector of the summed inverse-variance weighted difference between the pixels of overlapping images. We solve this equation using standard linear least-squares and then add the derived constant offsets to each CCD image. Next, we resample and coadd the data onto a common tangent plane and measure the median sky background (after aggressively masking sources) in an annulus defined to be $2-4$ times the \textit{masking radius} of each galaxy. We define an unique masking radius for each galaxy based on its approximate projected angular size (in arcseconds) on the sky, although our results are not sensitive to the exact choice of masking radius since our background annulus is well outside the outer, low surface-brightness envelope of the galaxy.

\subsection{Survey design and sample selection}
\label{sec:surveydesign}


To facilitate comparisons with simulations, we define a straightforward sample of isolated galaxies based only on luminosity and recessional velocity given by the HyperLeda database\footnote{http://leda.univ-lyon1.fr/}~\citep{HyperLeda}.
We consider galaxies with line-of-sight velocities in the Local Group rest frame ~\citep{LGRestFrame} in the range $2000< \mathrm{V}_{\rm LG} < 7000$ km s${^{-1}}$ ($ 30 \lesssim D  \lesssim 100  $ Mpc), with a $K$-band absolute magnitude M$_{K}< -19.6$ and within the footprints of the surveys described in Sec.~\ref{sec:legacy}.

The majority of $K$-band magnitudes are taken from the 2MASS survey~\citep{2MASS} and corrected for foreground extinction using reddening values from \citet{DustMap:SchlaflyFinkbeiner2011}, assuming $R_V = 3.1$.
No colour or morphological selections are imposed on our sample. 
We exclude targets that lie close to the Zone of Avoidance (Galactic latitude $|b| < 20$~deg) to avoid the area severely contaminated by Galactic dust and high stellar densities. 

To select `isolated' galaxies we use a simple criterion which requires that there is no neighbour brighter than $K_{\rm gal}=2.5$~mag with $|\Delta V|<250$ km s$^{-1}$ within a projected radius of 1~Mpc around each target. This criterion includes the brightest galaxies of `fossil groups' but excludes strongly interacting groups of galaxies with comparable masses. The criterion is based on a typical turn-around radius of 1~Mpc for the Local Group and other nearby groups such as M81 and Centaurus \citep{2009MNRAS.393.1265K}. The typical velocity dispersion of such groups is 70--80 km s$^{-1}$ \citep{2005AJ....129..178K}; we use a limit of 250 km s$^{-1}$ as a 3$\sigma$ threshold. Applying these criteria to the HyperLeda database yields a total sample of $\sim 3100$ target galaxies.

The luminosity range of our sample ensures that the target galaxies have a broad range of morphologies, star formation rates and stellar masses. This selection includes $\sim 940$  Milky-Way analogue systems, which are defined by their K-band luminosity ($-24.6<M_{K}<-23.0$) and the local environment criteria given in \citet{geha_2017}. Our lower distance bound of $\sim 30$~Mpc ensures that the virial radius of a typical galaxy in the sample ($\sim 250$--300~kpc) can be covered with a field of view (FOV) of $30\arcmin \times 30 \arcmin$ centred on a target (see Sec.~\ref{sec:cutout_method}). 
Furthermore, the image quality from the {\it DESI Legacy Imaging Surveys} is sufficiently high to discover stellar streams even at the upper distance bound of our survey, $\sim 100$~Mpc. Galaxies at that distance require only $10\arcmin \times 10 \arcmin$ cutouts. This outer boundary of the survey volume was chosen to obtain a sample containing a statistically significant number of tidal features based on the detection frequency in shallower surveys \citep{morales_2018}.

\begin{figure*}
\centering
	\includegraphics[width=0.85\textwidth]{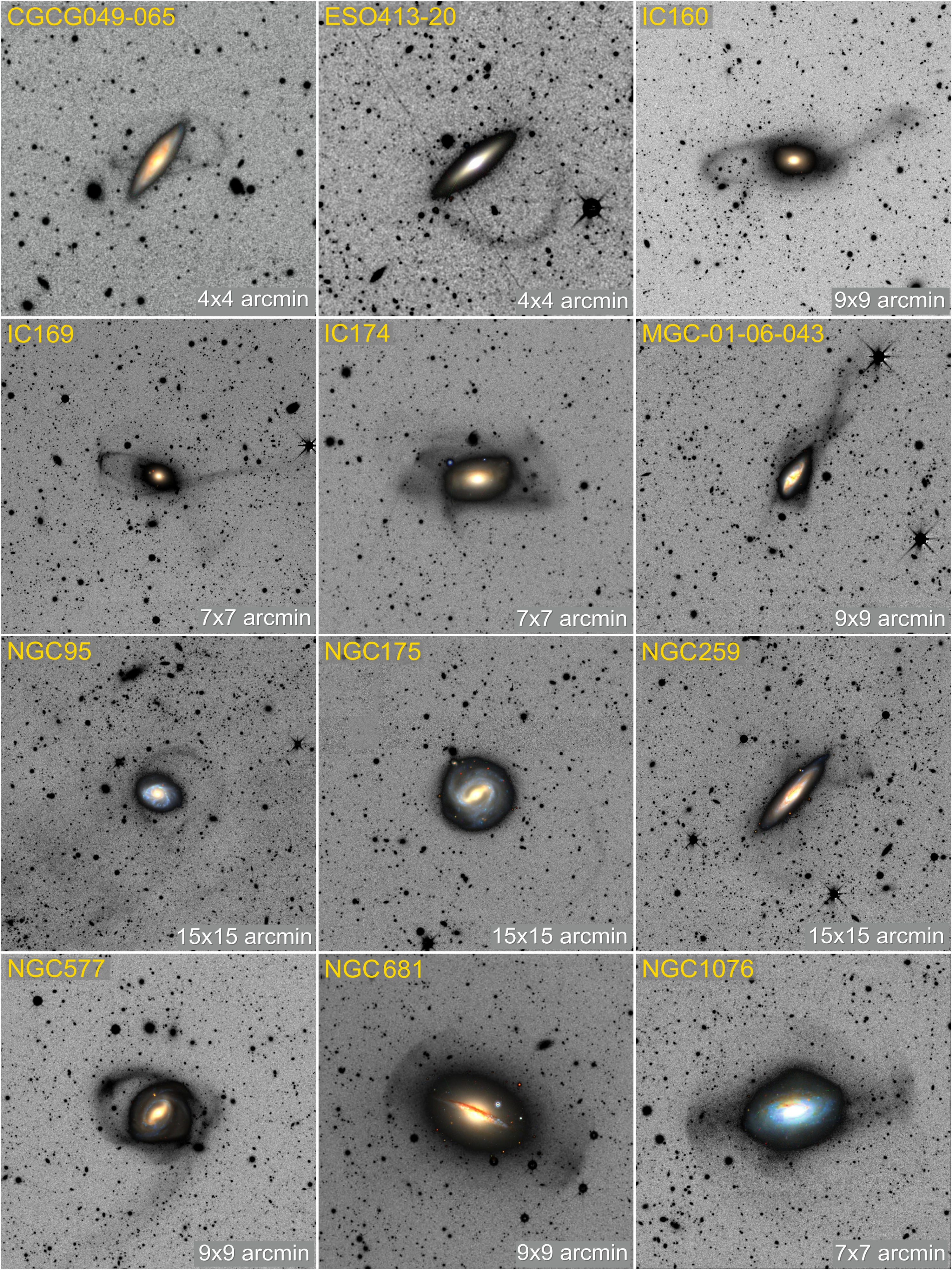}
        \caption{Stacked $g$ and $r$ image cutouts of 12 stellar streams discovered during the {\it proof-of-concept} phase of the {\it Stellar Stream Legacy Survey} introduced in this paper.
        For illustrative purposes, colour insets of the central region of the host galaxies have been added to the negative version of the images.}
    \label{fig:Fig3A_July31}
\end{figure*}

\begin{figure*}
    \centering
	\includegraphics[width=0.85\textwidth]{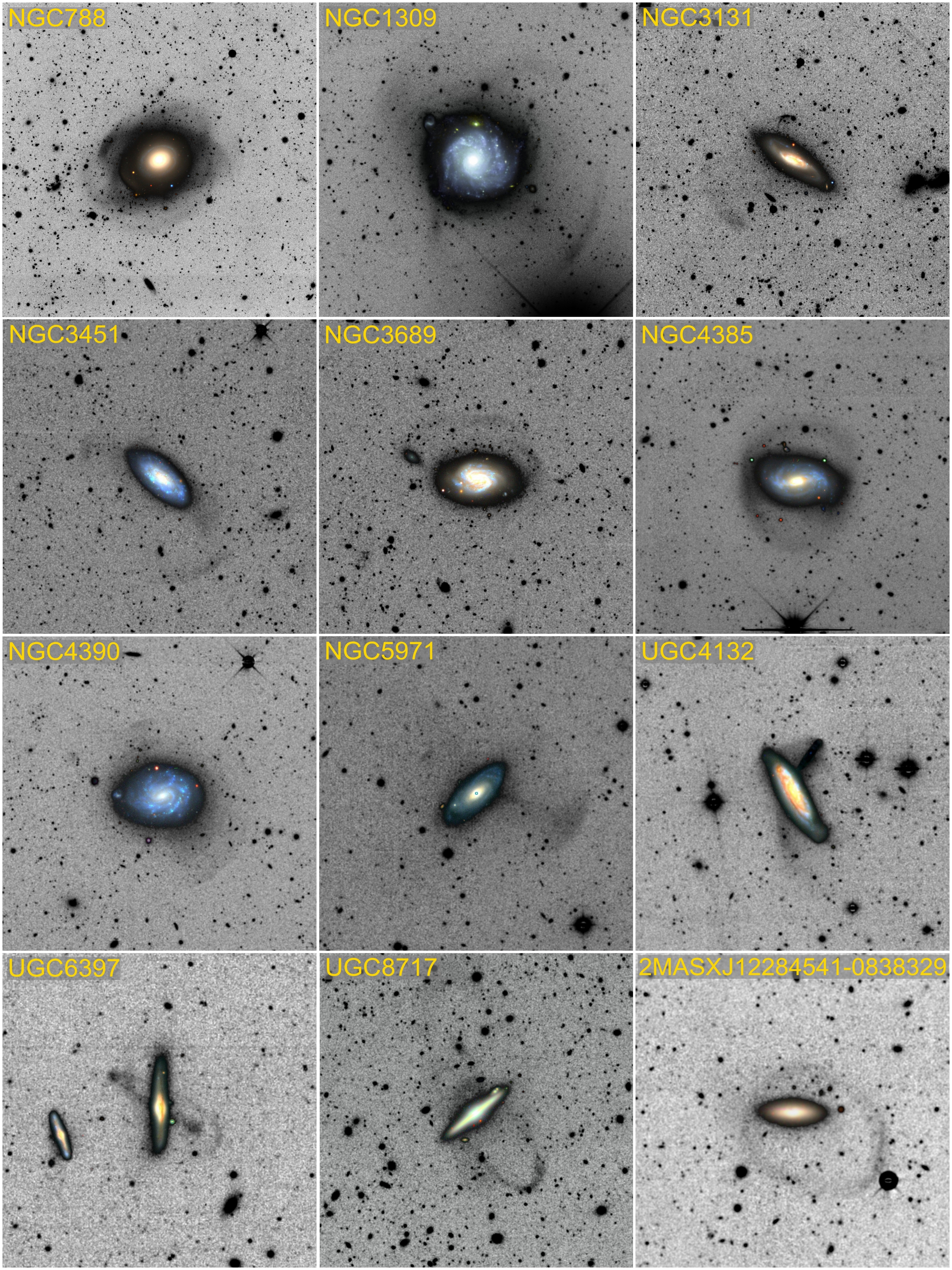}
        \caption{As Figure~\ref{fig:Fig3A_July31}, but showing an additional 12 new stellar streams from our {\it Stellar Stream Legacy Survey}.}
    \label{fig:Fig3B_July31}
\end{figure*}

\begin{figure*}
  \includegraphics[width=\textwidth]{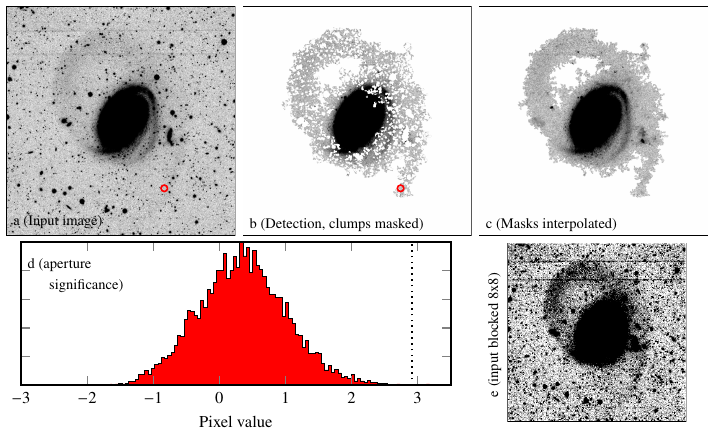}
  \caption{Detecting and measuring streams.
    (a) The input image:  NGC 4981 in $z$ band.
    The red aperture has a radius of $\detectionaperradarcsec$ arcsec and lies on the thinnest and most diffuse end of the stream.
    (b) All pixels not belonging to the galaxy (detected by \textsf{Noise\-Chisel}) or its stream are masked, including the pixels of objects that are in the background or foreground of the stream (´clumps´ in \textsf{Segment}).
    (c) All masked pixels interpolated by the median of ten neighbors.
    (d) The flux in the red aperture (dotted line), compared to a distribution of $\detectionapernumrandom$ similarly sized apertures, randomly placed over undetected pixels of the image ($\detectionaperupsigma\sigma$).
    (e) The input dataset warped to a pixel grid where each pixel is $8\times8$ larger than the input's pixels, to demonstrate the detection success on the input resolution.}  
  \label{fig-detection}
\end{figure*}

\begin{table*}
\centering
{\small
\caption{List of galaxies with stellar streams discovered during the proof-of-concept phase of our new survey. All the columns are based on homogenised information collected from the HyperLeda database. ({\it Column 1}): Galaxy name; ({\it column 2}): Morphological Hubble type of the galaxy; ({\it column 3}): Numerical code (T) in the de Vaucouleurs scale defined in RC2; ({\it column 4}): Apparent total $B$-band magnitude reduced to RC3 system;
({\it column }5): Apparent total $K$-band magnitude mostly based on 2MASS data; ({\it column 6}): Radial velocity with respect to the Local Group centroid \citep{LGRestFrame}; ({\it column 7}): Redshift distance $D$ assuming $H_0=73\,\mathrm{ km\,s^{-1}\,Mpc^{-1}}$; and ({\it column 8}): Field of view of the image cutouts showed in Figs.~\ref{fig:Fig3A_July31} and \ref{fig:Fig3B_July31}.}
\label{tab:target_list}

\begin{tabular}{llrrrr@{\,$\pm$\,}lrr@{\,$\times$\,}l}
\hline\hline
Name & Type & 
\multicolumn{1}{c}{T} & 
\multicolumn{1}{c}{$B_T$} & 
\multicolumn{1}{c}{$K_T$} & 
\multicolumn{2}{c}{V$_{\rm LG}$} & 
\multicolumn{1}{c}{$D$} & 
\multicolumn{2}{c}{FOV} \\
 & & & 
\multicolumn{1}{c}{mag} & 
\multicolumn{1}{c}{mag} & 
\multicolumn{2}{c}{km\,s$^{-1}$} & 
\multicolumn{1}{c}{Mpc} & 
\multicolumn{2}{c}{arcmin} \\

\hline
2MASXJ12284541-0838329  & S0	&$ -2.0 \pm 1.7 $&$ 15.59 \pm 0.50 $&$ 11.19 \pm 0.08 $&  5245 & 39 &  71.8 & 7 & 7 \\
CGCG 049--065				& S?	&$  1.8 \pm 4.8 $&$ 15.63 \pm 0.28 $&$ 11.44 \pm 0.05 $& 11222 & 5  & 153.7 & 4 & 4 \\
ESO 413--020				& S0	&$ -2.0 \pm 0.4 $&$ 14.30 \pm 0.47 $&$ 10.63 \pm 0.06 $&  6011 & 36 &  82.3 & 4 & 4 \\
IC 0160						& E-S0	&$ -2.8 \pm 0.7 $&$ 14.88 \pm 0.39 $&$ 10.36 \pm 0.09 $&  5611 & 24 &  76.8 & 9 & 9 \\
IC 0169						& E-S0	&$ -3.0 \pm 2.4 $&$ 15.80 \pm 0.49 $&$ 11.09 \pm 0.08 $&  8522 & 25 & 116.7 & 7 & 7 \\
IC 0174						& S0	&$ -1.9 \pm 0.5 $&$ 14.52 \pm 0.32 $&$ 10.55 \pm 0.03 $&  5346 & 14 &  73.2 & 7 & 7 \\
MCG $-$01--06--043			& Sbc	&$  4.0 \pm 1.3 $&$ 14.80 \pm 0.42 $&$ 10.46 \pm 0.04 $&  5110 & 20 &  70.0 & 9 & 9 \\
NGC 0095					& SABc	&$  5.2 \pm 0.5 $&$ 13.28 \pm 0.04 $&$  9.48 \pm 0.04 $&  5584 & 2  &  76.5 &15 & 15\\
NGC 0175					& Sab	&$  2.2 \pm 0.5 $&$ 12.98 \pm 0.06 $&$  9.23 \pm 0.06 $&  3949 & 17 &  54.1 &15 & 15\\
NGC 0259					& Sbc	&$  4.0 \pm 0.6 $&$ 13.70 \pm 0.41 $&$  9.22 \pm 0.05 $&  4182 & 5  &  57.3 &15 & 15\\
NGC 0577					& Sa	&$  1.0 \pm 0.4 $&$ 14.12 \pm 0.32 $&$  9.91 \pm 0.08 $&  6048 & 4  &  82.8 & 9 & 9 \\
NGC 0681					& SABa	&$  2.0 \pm 0.7 $&$ 12.77 \pm 0.07 $&$  8.67 \pm 0.03 $&  1805 & 3  &  24.7 & 9 & 9 \\
NGC 0788                    & S0-a  &$  0.0 \pm 0.5 $&$ 13.06 \pm 0.06 $&$  9.10 \pm 0.05 $&  4144 & 24 &  60.8 &10 &10 \\
NGC 1076					& S0-a	&$  0.2 \pm 1.3 $&$ 13.71 \pm 0.50 $&$ 10.00 \pm 0.04 $&  2084 & 4  &  28.5 & 7 & 7 \\
NGC 1309					& Sbc	&$  3.9 \pm 0.6 $&$ 11.96 \pm 0.11 $&$  9.12 \pm 0.04 $&  2092 & 2  &  28.6 & 7 & 7 \\
NGC 3131					& Sb	&$  3.1 \pm 0.5 $&$ 13.88 \pm 0.28 $&$  9.69 \pm 0.03 $&  4974 & 2  &  68.1 & 7 & 7 \\
NGC 3451					& Sc	&$  6.5 \pm 0.9 $&$ 13.50 \pm 0.31 $&$ 10.26 \pm 0.07 $&  1260 & 2  &  17.2 & 7 & 7 \\
NGC 3689					& SABc	&$  5.3 \pm 0.5 $&$ 13.01 \pm 0.11 $&$  9.20 \pm 0.03 $&  2667 & 2  &  36.5 & 7 & 7 \\
NGC 4385					& S0-a	&$ -0.1 \pm 2.1 $&$ 13.15 \pm 0.06 $&$  9.69 \pm 0.37 $&  1988 & 3  &  27.2 & 7 & 7 \\
NGC 4390					& Sc	&$  5.0 \pm 0.8 $&$ 13.34 \pm 0.03 $&$ 10.54 \pm 0.25 $&   983 & 3  &  13.5 & 7 & 7 \\
NGC 5971					& Sa	&$  1.1 \pm 0.5 $&$ 14.72 \pm 0.32 $&$ 11.08 \pm 0.12 $&  3555 & 2  &  48.7 & 9 & 9 \\
UGC 04132					& Sbc	&$  4.0 \pm 0.3 $&$ 13.84 \pm 0.51 $&$  9.63 \pm 0.04 $&  5160 & 5  &  70.7 & 7 & 7 \\
UGC 06397					& Sab	&$  2.0 \pm 0.2 $&$ 14.90 \pm 0.29 $&$ 10.34 \pm 0.06 $&  6231 & 39 &  85.3 & 7 & 7 \\
UGC 08717					& SABa	&$  1.0 \pm 0.5 $&$ 14.67 \pm 0.29 $&$ 10.05 \pm 0.05 $&  4686 & 3  &  64.2 & 9 & 9 \\
\hline\hline
\end{tabular}

}
\end{table*}

\begin{figure}
\begin{center}
  \includegraphics[width=0.8\columnwidth]{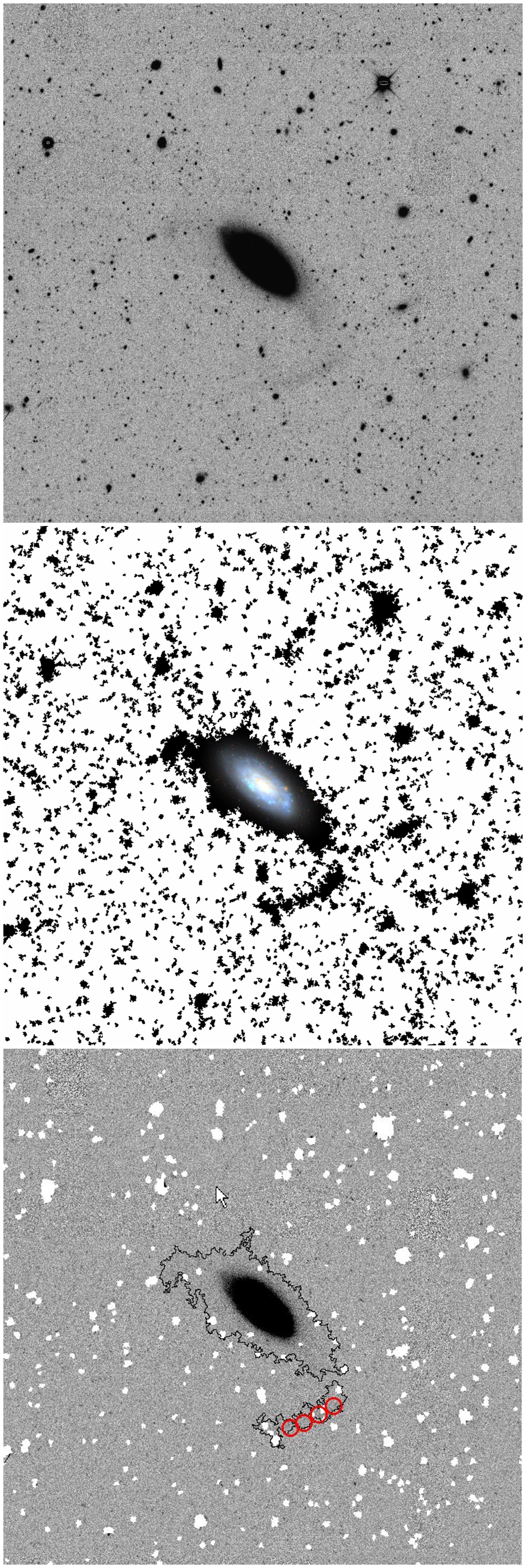}
  \end{center}

  \caption{Example of our aperture photometry approach applied to the stellar stream detected around NGC 3451.
    ({\it Top panel}): Input image in $g$-band, with subtracted sky, showing the stream around it. ({\it Middle panel}): The detection contour of the galaxy and its stream are highlighted ({\it black line}) on a pixel grid $8\times8$ courser than the input image to highlight the low surface brightness signal. ({\it Bottom panel}): The apertures used to measured the surface brightness, magnitude and colours are placed along the stream ({\it red circles}). The foreground stars and background sources are masked ({\it white areas}).}  
  \label{fig-photometry}
\end{figure}

\section{Data analysis and photometry}
\label{sec:analysis}


 To date, we have visually inspected a total of 689 cutouts from all the targets included in the DES sky footprint, yielding a sample of 89 candidate streams. Some additional conspicuous stellar streams have also found in an incomplete visual inspection of the target cutouts from the DECaLs and MsLS sky footprints.
In this paper, as a proof of concept, we present a quantitative study of the detection significance and photometric properties for a representative sub-sample of 24 of those candidates. Images of this sub-sample are shown in Figs.~\ref{fig:Fig3A_July31} and \ref{fig:Fig3B_July31}, and the properties of  their host galaxies are listed in Table \ref{tab:target_list}. These features are reported for the first time here. The sub-sample was chosen to span a representative range of morphological types, surface brightnesses and distances in the DES and DECaLs sky footprints. Our aim in the following analysis is to asses the detection limits of our survey and the viability of our baseline photometric approach. A catalogue of candidates for the complete sample of hosts are published in a forthcoming paper (Mir\'o-Carretero et al., in preparation).

\subsection{Detection and photometry of stellar tidal streams}
\label{sec:photometry}


 Our strategy for searching for stellar streams is mainly based on visual inspection of DESI LS imaging of the outskirts of the target galaxies.  Streams are inherently non-symmetric and elongated, making them extremely hard to distinguish from the noise with classical detection methods. Therefore, we use \textsf{Noise\-Chisel} \citep{Akhlaghi15}, a component of the GNU Astronomy Utilities \detectiongnuastrover{} (Gnuastro\footnote{\url{http://www.gnu.org/software/gnuastro}}),  to confirm detections and estimate the significance of each detected structure in our survey.\footnote{Kado-Fong et al. (2018) proposed a method to automatically detect which galaxies have streams (or any kind of non-smooth faint structure) based on successive convolution of the images with the same kernel. Although this approach is robust, their suggested kernel and parameters are specific to their study of a more distant galaxy sample at $0.05<z<0.45$ with Subaru Hyper Suprime-Cam imaging. We have tried to apply such a solution for the relatively massive, closer galaxies in our sample, without success. We find that this approach is not able to remove the host galaxy as clearly as they showed in their images, mainly because our target galaxies have a much larger sizes (in pixels) than the small kernel suggested by Kado-Fong et al. (2018). Further investigation into a suitable form of the kernel are needed to optimise this approach for samples of nearby galaxies such as ours.}

\textsf{Noise\-Chisel} uses thresholds that are below the sky level to detect extremely faint signals without making assumptions about the shape or surface brightness profile of the sources.
It then uses simple mathematical morphology operators (erosion and dilation) to extract a contiguous signal at low surface brightness.
For example \citet{Akhlaghi20} describe a successful application to detect the outer wings of M51 to a surface brightness of 25.97 $\mathrm{mag\,arcsec^{-2}}$ ($r$-band) in a single $\sim1$ minute SDSS exposure.

Figure \ref{fig-detection}(a) shows an example of NGC 4981 in the $z$-band from our dataset\footnote{Because the focus here is on the large and extended streams, \textsf{Noise\-Chisel} was given a Gaussian kernel of FWHM$=\detectionnckernelfwhm$ pixels (default 2 pixels), truncated at $\detectionnckerneltrunc$ times the FWHM. As well as \texttt{-{}-tilesize=\detectionnctilesize} and \texttt{-{}-detgrowmaxholesize=\detectionncholesize}, the default values were used for all other parameters.}.
All pixels that do not belong to NGC 4981 and its stream have been masked in panel (b), as well as all foreground and/or background sources overlapping the stream.
Line-of-sight contamination was identified as clumps over \textsf{Noise\-Chisel}'s detections, using \textsf{Segment}\footnote{To better separate the extended emission of background galaxies or foreground stars as clumps (to be masked), \textsf{Segment} was run with a Gaussian kernel of FWHM$=\detectionnckernelfwhmseg$ (default 1.5 pixels), the default values were used for all other parameters.} \citep[also part of Gnuastro]{Akhlaghi20} and masked in Figure \ref{fig-detection}(b).
Figure \ref{fig-detection}(c) shows the host and the stream with all masked clumps interpolated.

The $\detectionaperradarcsec$ arcsec radius aperture shown in (a) is positioned along the thinnest and faintest end of the detected stream at a point with no background or foreground source.
The flux within this aperture has a surface brightness of $\detectionapersb$ $\mathrm{mag\,arcsec^{-2}}$, which has a significance of $\detectionaperupsigma\sigma$ compared to $\detectionapernumrandom$ random apertures of the same size placed over undetected regions (see panel (d)).
As the area increases, the significance of the detection improves, enabling us to extract meaningful colour measurements.
However, we are not limited to measuring colours in simple aperture geometries.
Because of \textit{Gnuastro}'s modularity (which separates image labelling, through detection or segmentation, from measurements and catalogue production), magnitudes and colours can be measured on any feature with arbitrary morphology, like the stellar streams \citep{Akhlaghi19}.

Another factor to consider is that the detectability of these
features is closely related to the surface brightness limit of
the image cutouts obtained in Sec.~\ref{sec:cutout_method}. This
limit depends on the region of sky in which the galaxy is
located, and in general on the exposure time and observing conditions. 
In this survey, the surface brightness limit of the images was calculated for the three bands following the standard method in Appendix A of \cite{2020A&A...644A..42R}, as implemented in Gnuastro's \textsf{MakeCatalog}\footnote{\url{https://www.gnu.org/software/gnuastro/manual/html_node/Surface-brightness-limit-of-image.html}}, that is the measured $3\sigma$ value in an area of 100 arcsec$^2$. 



A high degree of accuracy in data reduction and post-processing is required to measure surface brightness, colours and integrated luminosities for extremely faint and diffuse features. The procedures detailed in previous sections provide a necessary improvement in the stability of the sky background estimate. However, other systematic uncertainties still need to be addressed. For example, some tidal features 
overlap with their host galaxy along the line of sight. Deblending these features to measure directly their total luminosity requires a precise model of the host galaxy. This is a challenging problem which we leave for future work.

We carried out a photometric analysis on the set of streams identified in our proof-of-concept sample by using {\it Gunastro}'s \textsf{MakeCatalog} on the sky-subtracted image generated by \textsf{NoiseChisel} \citep{Akhlaghi15}. The surface brightness of each stream was measured in the $r$, $g$ and $z$ bands, together with the colours $(g-r)$ and $(r-z)$. The measurements were performed using circular apertures placed on the detected track of the stream, after all foreground and background sources were removed. Regions where the stream surface brightness could be contaminated by the host galaxy were avoided. This method is illustrated in Figure \ref{fig-photometry}, which shows an example application to a tidal stream around NGC3451. 
The average surface brightness of this stream is only 1.24 {$\mathrm{mag\,arcsec^{-2}}$} brighter than the surface brightness limit of the image. 

The resulting average surface brightnesses and colours of the stellar streams from our proof-of-concept sample are given in Table~\ref{tab:streams_photo} along with the corresponding errors, computed by Gnuastro's \textsf{MakeCatalog}\footnote{\url{https://www.gnu.org/software/gnuastro/manual/html_node/Magnitude-measurement-error-of-each-detection.html}}. This table also includes the surface brightness limit for the $r$ band, which is representative of the depth of the corresponding images.

In order to quantify the significance of individual stream detections, we define the {\it Detection Index}, $DI_\mathrm{stream} = (F_\mathrm{stream} - F_\mathrm{blank}) / \sigma$. We note that $F_\mathrm{stream}$ is the flux measured in a single aperture on the stream, $F_\mathrm{blank}$ is the median flux measured in 10,000 apertures of the same size randomly placed in regions of the sky-subtracted image with no detection\footnote{This is the quantity referred to as the 'Upper Limit' in \textit{Gnuastro}, see \url{https://www.gnu.org/software/gnuastro/manual/html_node/Upper-limit-magnitude-of-each-detection.html}}, and $\sigma$ is the standard deviation of the same 10,000 background flux measurements. Columns 3 and 4 in Table~\ref{tab:streams_photo} report the maximum and mean value of $DI_\mathrm{stream}$, respectively, taken over all the apertures placed on each stream.

\begin{figure}
    \includegraphics[width=0.85\columnwidth]{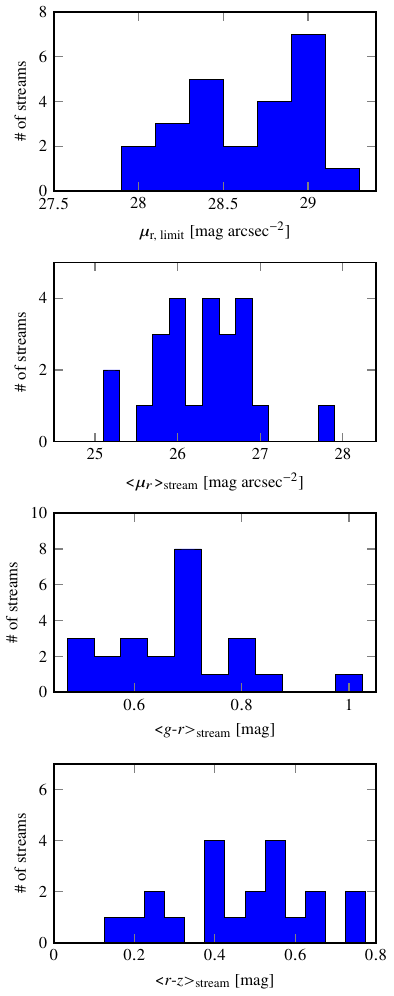}
    \caption{Histograms showing the distribution of the depth of the image cutouts and of the photometry measurements of the stellar streams around the galaxies listed in Table~\ref{tab:streams_photo}. From top to bottom: Surface brightness limit for the $r$ band, surface brightness for the $r$ band, and colours $(g-r)$ and $(r-z)$.}
    \label{fig-streamphoto}
\end{figure}

 Fig.~\ref{fig-streamphoto} shows distributions of the surface brightness limit of our images and the surface brightnesses and colours of the stellar streams identified around the galaxies listed in Table \ref{tab:streams_photo}. Taking the colour distributions in Fig.~\ref{fig-streamphoto} at face value, it appears that the $g-r$ colours of streams are systematically redder than those of typical dwarf satellite galaxies. For example, the satellite galaxies around Milky Way analogues observed in the SAGA survey\footnote{The SAGA colour-magnitude selection excludes galaxies with $g-r \gtrsim 0.7$ at $r \gtrsim 17$.} have a range $0.2 < g-r < 0.7$ \citep[][]{mao_2021}, whereas we find stream colours $0.5 < g-r < 0.8$ around visually similar hosts. Under the simplistic assumption that the colours measured in SAGA are representative of satellite dwarf galaxies up to the point which their star formation is truncated by tidal disruption, after which they redden, our result may imply that the stream population is, on average, dynamically older than surviving luminous satellites. Redder colours may also imply that progenitors of the streams we detect in our survey are more massive than typical surviving satellites. Both interpretations would be consistent with theoretical expectations \citep[e.g.][]{cooper_2010}. However, a robust statistical analysis requires both the larger sample of our complete survey and a careful exploration of the systematic uncertainties on our colour measurements, in addition to a more detailed study of model predictions.



\begin{table*}
\centering
{\small
\caption{Photometry of stellar streams discovered in our proof-of-concept analysis for a sub-sample of {\it Stellar Stream Legacy Survey} targets, listed in Tab.~\ref{tab:target_list}. From left to right, columns 2--4 give surface brightness limits in the $r$ band and the `Detection Index', as defined in Section \ref{sec:photometry}. Columns 5 to 9 show the surface brightness in all three bands and the colours of the streams averaged over all the apertures placed on the stream. $^\dagger$The designation of the galaxy in the first row is abbreviated from 2MASXJ12284541-0838329.}
\label{tab:streams_photo}
\begin{tabular}{lcccccccc}

Host galaxy & $\mu_\mathrm{r, limit}$ & \multicolumn{2}{c}{$\mathrm{DI}_\mathrm{stream}$} & $\langle \mu_{g}\rangle_\textrm{stream}$ & $\langle\mu_{r}\rangle_\mathrm{stream}$ & $\langle\mu_{z}\rangle_\mathrm{stream}$ & $\langle g - r \rangle_\mathrm{stream}$ & $\langle r-z \rangle_\mathrm{stream}$  \\

 & & maximum & average & & & & & \\
 
  & [mag arcsec$^{-2}$]  & $\sigma$ & $\sigma$ & [mag arcsec$^{-2}$] & [mag arcsec$^{-2}$] & [mag arcsec$^{-2}$] & [mag] & [mag] \\
\hline\hline

\small 2MASXJ12284541$^{\dagger}$\normalsize &27.89&17.38   &11.08   &26.67 $\pm$  0.07&25.98 $\pm$  0.08&25.47 $\pm$  0.11&0.69 $\pm$  0.11&0.50 $\pm$  0.14 \\
CGCG049-065            &28.09&12.84   &6.62    &27.16 $\pm$  0.20&26.46 $\pm$  0.15&26.07 $\pm$  0.23&0.70 $\pm$  0.25&0.39 $\pm$  0.28 \\
ESO413-20              &28.78&13.07   &10.17   &27.44 $\pm$  0.10&26.76 $\pm$  0.08&26.55 $\pm$  0.18&0.68 $\pm$  0.13&0.21 $\pm$  0.20 \\
IC0160                 &28.73&47.78   &33.66   &26.26 $\pm$  0.04&25.60 $\pm$  0.03&25.10 $\pm$  0.05&0.67 $\pm$  0.05&0.50 $\pm$  0.06 \\
IC0169                 &29.02&37.16   &15.98   &26.83 $\pm$  0.04&26.32 $\pm$  0.04&25.79 $\pm$  0.06&0.51 $\pm$  0.05&0.53 $\pm$  0.07 \\
IC0174                 &28.95&47.21   &32.62   &26.26 $\pm$  0.03&25.59 $\pm$  0.02&25.16 $\pm$  0.04&0.67 $\pm$  0.03&0.44 $\pm$  0.04 \\
MGC-01-06-043          &28.99&30.41   &26.39   &26.63 $\pm$  0.03&25.99 $\pm$  0.02&25.46 $\pm$  0.04&0.64 $\pm$  0.04&0.54 $\pm$  0.05 \\
NGC0095                &28.34&11.66   &6.93   &27.22 $\pm$  0.04&26.77 $\pm$  0.05&26.36 $\pm$  0.08&0.46 $\pm$  0.06&0.40 $\pm$  0.09 \\
NGC0175                &28.34&6.07    &4.05    &28.36 $\pm$  0.21&27.62 $\pm$  0.16&26.87 $\pm$  0.18&0.74 $\pm$  0.27&0.75 $\pm$  0.24 \\
NGC0259                &28.84&24.11   &19.26   &27.00 $\pm$  0.04&26.32 $\pm$  0.03&25.79 $\pm$  0.06&0.68 $\pm$  0.05&0.53 $\pm$  0.07 \\
NGC0577                &28.86&78.57   &58.96   &25.78 $\pm$  0.03&25.13 $\pm$  0.02&24.63 $\pm$  0.03&0.66 $\pm$  0.03&0.50 $\pm$  0.04 \\
NGC0681                &28.90&14.49   &11.60   &25.94 $\pm$  0.02&25.18 $\pm$  0.02&24.78 $\pm$  0.02&0.76 $\pm$  0.03&0.40 $\pm$  0.03 \\
NGC0788                &28.87&48.65   &27.23   &26.81 $\pm$  0.05&25.98 $\pm$  0.03&25.41 $\pm$  0.06&0.83 $\pm$  0.05&0.57 $\pm$  0.06 \\
NGC1076                &28.87&43.79   &35.08   &26.48 $\pm$  0.02&25.95 $\pm$  0.02&25.55 $\pm$  0.04&0.53 $\pm$  0.03&0.40 $\pm$  0.04 \\
NGC1309                &28.76&24.42   &23.02   &26.26 $\pm$  0.02&25.66 $\pm$  0.02&25.39 $\pm$  0.03&0.61 $\pm$  0.02&0.27 $\pm$  0.04 \\
NGC3131                &28.23&16.29   &13.36   &27.25 $\pm$  0.06&26.57 $\pm$  0.05&26.35 $\pm$  0.12&0.67 $\pm$  0.08&0.22 $\pm$  0.13 \\
NGC3451                &27.93&8.09    &7.10    &27.20 $\pm$  0.08&26.69 $\pm$  0.09&25.96 $\pm$  0.09&0.50 $\pm$  0.12&0.73 $\pm$  0.13 \\
NGC3689                &28.00&10.75   &6.45    &27.73 $\pm$  0.16&26.93 $\pm$  0.13&26.80 $\pm$  0.30&0.80 $\pm$  0.21&0.13 $\pm$  0.33 \\
NGC4385                &28.44&20.49   &10.69   &26.68 $\pm$  0.05&26.20 $\pm$  0.04&25.58 $\pm$  0.07&0.49 $\pm$  0.06&0.61 $\pm$  0.08 \\
NGC4390                &28.10&11.91   &9.01    &27.31 $\pm$  0.06&26.73 $\pm$  0.07&26.56 $\pm$  0.12&0.58 $\pm$  0.09&0.17 $\pm$  0.13 \\
NGC5971                &28.50&21.84   &18.93   &26.79 $\pm$  0.03&26.01 $\pm$  0.02 & unreliable     &0.78 $\pm$  0.04 & unreliable \\
UGC04132               &28.73&17.45   &8.28    &27.27 $\pm$  0.11&26.31 $\pm$  0.07 & unreliable     &0.96 $\pm$  0.12 & unreliable \\
UGC06397               &28.29&25.78   &14.34   &26.36 $\pm$  0.04&25.76 $\pm$  0.05 & unreliable     &0.60 $\pm$  0.06 & unreliable \\
UGC08717               &28.20&12.70   &8.93    &27.05 $\pm$  0.07&26.44 $\pm$  0.08&25.83 $\pm$  0.10&0.60 $\pm$  0.11&0.61 $\pm$  0.13 \\

\hline\hline
\end{tabular}
}
\end{table*}


\subsection{Distinguishing stellar streams from other features}
\label{sec:confusion}

As stated in Sec.\,\ref{sec:intro}, our aim is to study dwarf galaxy accretion around nearby galaxies that have not been strongly perturbed by major merger events in the recent past. Therefore we need to develop a strategy to identify and exclude from our sample 
features
associated with the partially relaxed debris of major mergers and 
perturbations to a central stellar disk that are visually similar to stellar streams.

\subsubsection{Old major-merger remnant systems}

Our isolation criterion should exclude the majority of massive galaxies and coalescence phases of major mergers. It should also remove from our sample very bright galaxies undergoing tidal disruption within massive galaxy clusters. However, galaxies in the late stages of major mergers may still enter our parent sample. Such `post-merger' systems may have a well-relaxed core surrounded by a large number of faint remnant features generated by past tidal interactions between the progenitors\footnote{Although they are not our primary target, features associated with low mass ratio mergers may also provide useful constraints on galaxy assembly. Our survey is likely to discover a significant number of such features.}.

It may be possible to disentangle these cases by exploiting the morphology of their tidal tails, in particular their width. The width of tidal features scales with the progenitor mass, $m$, as $m^{1/3}$ \citep[e.g.][]{johnston_2001,erkal_2016}. Older, more massive mergers have significantly broader features. A similar distinction was used in \cite{johnston_2001} to classify debris. 

\begin{figure*}
 \includegraphics[width=1.0\textwidth]{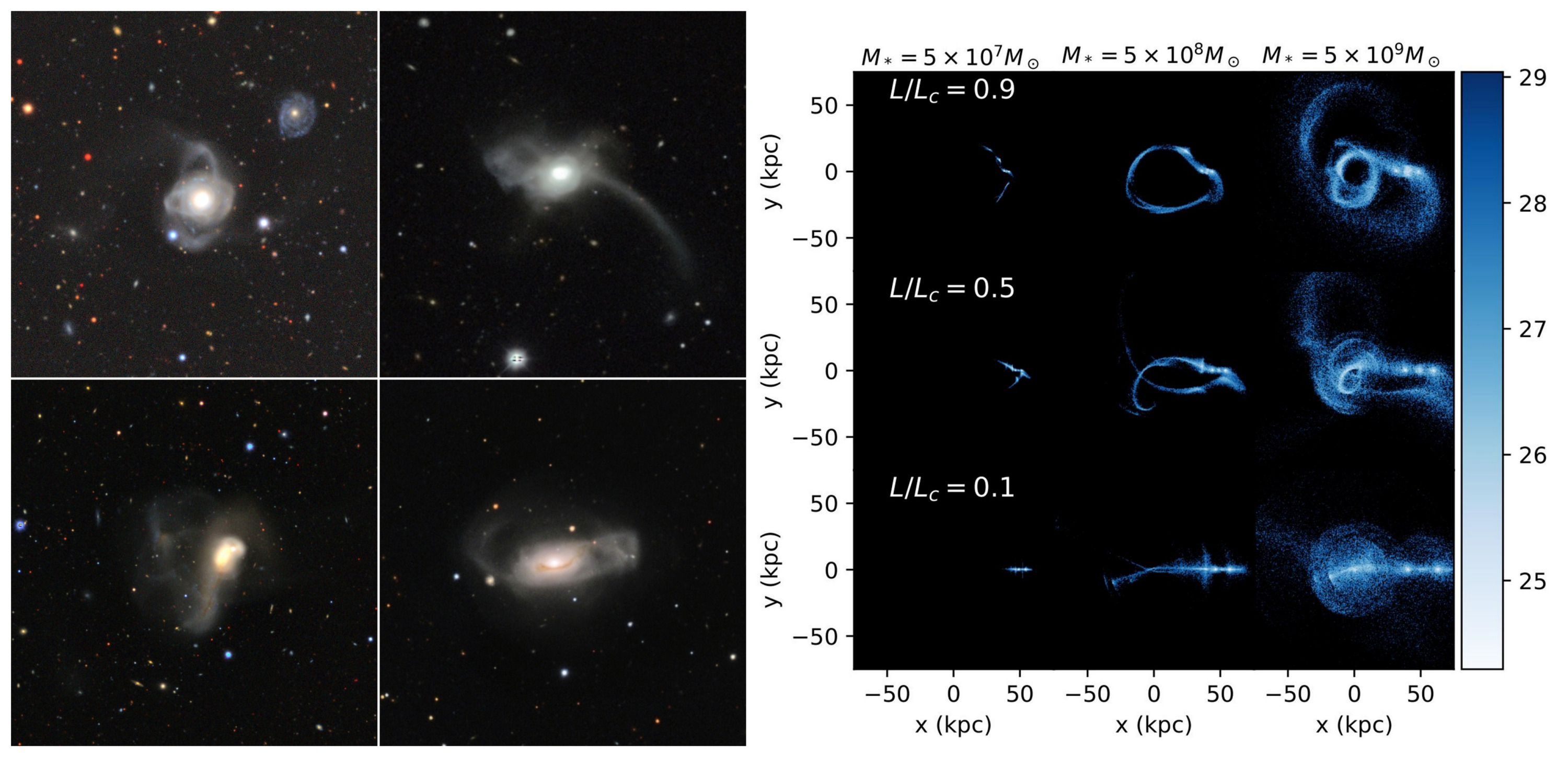}
          \caption{({\it Left panels}): Examples of likely major merger remnants in our  proof-of-concept sample (see Table ~\ref{tab:target_list}): ESO474-26, NGC 655,
          NGC 3303 and NGC 7288. These were identified as massive merger remnants based on the similarity to the outcome of idealised simulations: see right panels. ({\it Right panels}):  Simulation snapshots of galaxy mergers evolved for 2 Gyr (see Sec. 4.1). Each corresponds to different initial conditions: rows show different orbital eccentricities given by their circularity $L/L_{c}$ as defined by \cite{2015MNRAS.454.2472H}; columns show different progenitor stellar masses. The $5\times 10^9 M_\odot$ progenitor corresponds to a major merger with a mass ratio $\sim$1:3; the other two models are for minor mergers. Progenitors were initialised on the size--luminosity relation of Local Group satellites \citep{2012AJ....144....4M}. The colour bar shows the surface brightness of the tidal debris in $\mathrm{mag\, arcsec^{-2}}$. 
          }
    \label{fig:stream_spray}
\end{figure*}

We demonstrate how this classification would work by running simulations of the tidal debris from different mass progenitors on a range of orbits. These simulations were carried out with the modified Lagrange Cloud stripping technique from \cite{2014MNRAS.445.3788G} as implemented in \cite{2017MNRAS.470...60E}. This technique faithfully reproduces the debris of an $N$-body disruption by ejecting particles from the Lagrange points of the progenitor, allowing us to model the disruption in a few CPU-seconds instead of a few CPU-hours. The progenitors are modelled as Plummer spheres following the size--luminosity relation for observed local group dwarfs \citep{brasseur2011}. The host galaxy potential was modelled using the \texttt{MWPotential2014} from \cite{2015ApJS..216...29B} which provides a good match to the Milky Way potential. For computational simplicity, we replace the bulge with a Hernquist profile \citep{1990ApJ...356..359H} with a mass of $5\times10^9 M_\odot$ and a scale radius of 0.5 kpc. 

Examples from our test simulations are shown in the right panel of Fig.~\ref{fig:stream_spray}. It is clear that more massive progenitors create broader tidal debris streams. In addition, as the orbit is made more radial, the debris transitions from being stream-like to being shell-like \citep[e.g.][]{2015MNRAS.454.2472H}. The highest mass considered here corresponds to a major merger with a total mass ratio of $\sim1:3$. We have used abundance matching for the stellar mass-halo mass relation \citep[e.g.][]{2013ApJ...770...57B,2013MNRAS.428.3121M}, and evolved the tidal debris for 2 Gyr. 

The simulations suggest a correlation between stream width, halo mass and dynamical age. We find the thinnest and most coherent structures are generated on the early phases of the interaction, when the central galaxy shows the strongest morphological distortions. This phase is also expected to generate starbursts and AGN activity that clearly distinguish merging systems \citep{Springel2005,Lotz2008,Scott2014,Pearson2019}. The debris is broader, and broadens more rapidly, than in a minor merger. In minor mergers, by definition, the central potential is not significantly perturbed, and the central galaxy remains almost quiescent with a small star formation enhancement and no AGN activation \citep{Gordon2019}. For comparison, the left-hand panel of Fig.~\ref{fig:stream_spray} shows  likely major merger remnants identified in our proof-of-concept sample. Quantitative comparison of a larger set of simulations and observations of partly-relaxed mergers is useful to develop more robust estimators of progenitor mass.



\begin{figure*}
	\includegraphics[width=1.0\textwidth]{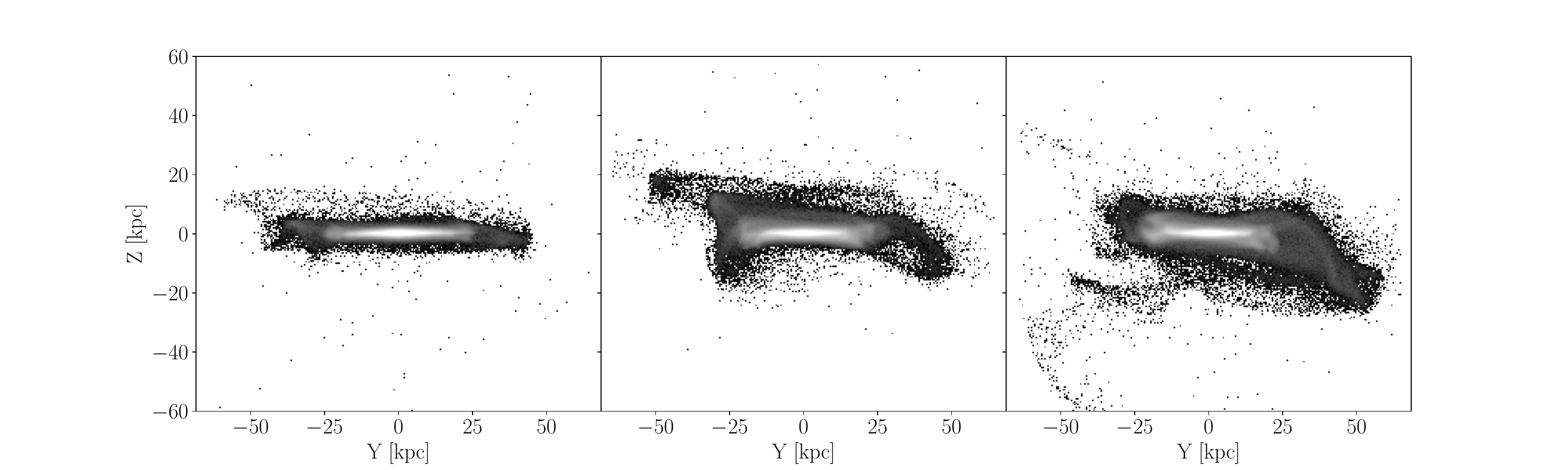}
	\includegraphics[width=1.0\textwidth]{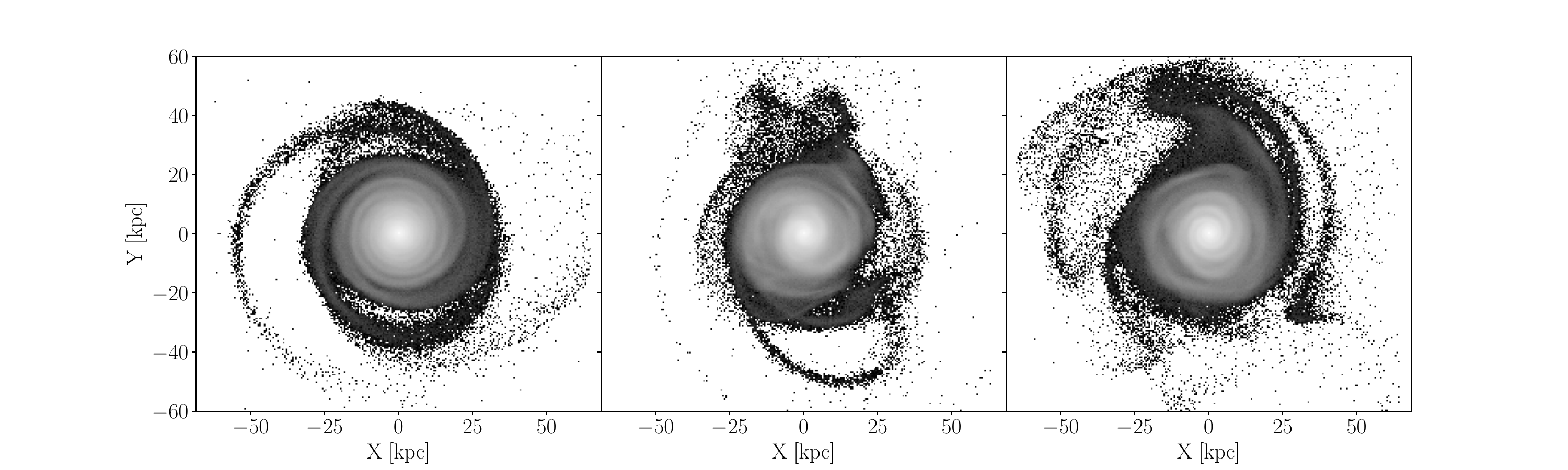}
	\includegraphics[width=1.0\textwidth]{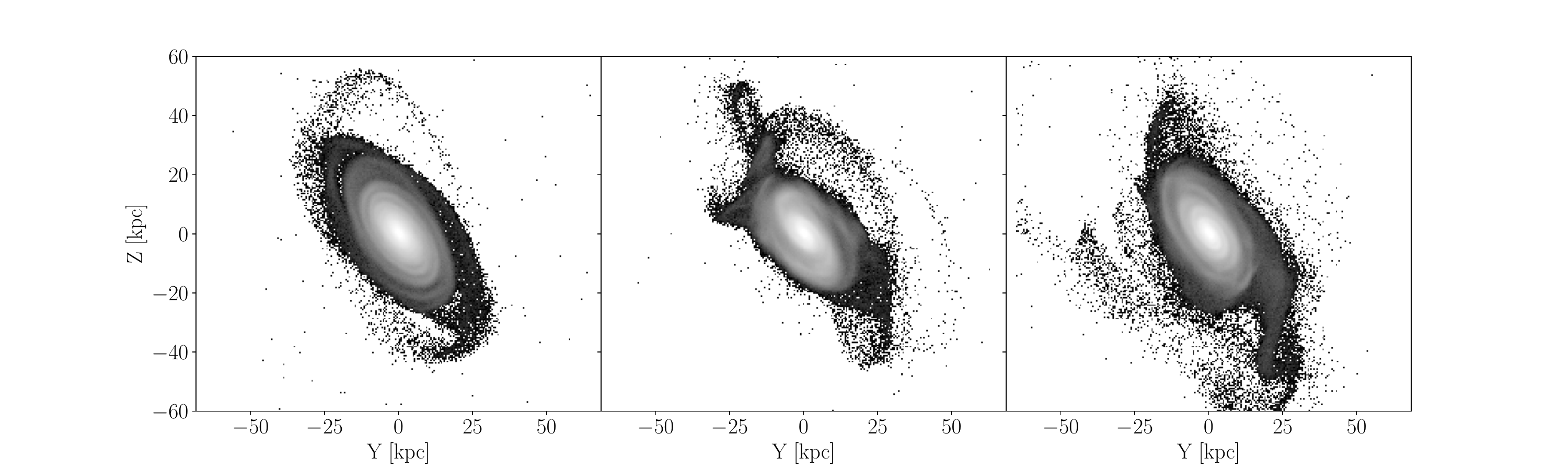}
	\caption{Gallery of simulated galactic feathers shown at different stages of interaction (different columns) and different viewing angles (rows). {\it Top:} edge-on view; {\it middle:} Face-on view; and {bottom:} Inclined view. Many of these structures show morphologies reminiscent of tidal streams from disrupting satellites although in this case this material is purely in-situ from the disc. A colour characterisation or metallicity measurement could potentially disentangle the two possibilities.}
    \label{fig-cfpl}
\end{figure*}
\begin{figure*}
    \centering
	\includegraphics[width=0.95 \textwidth]{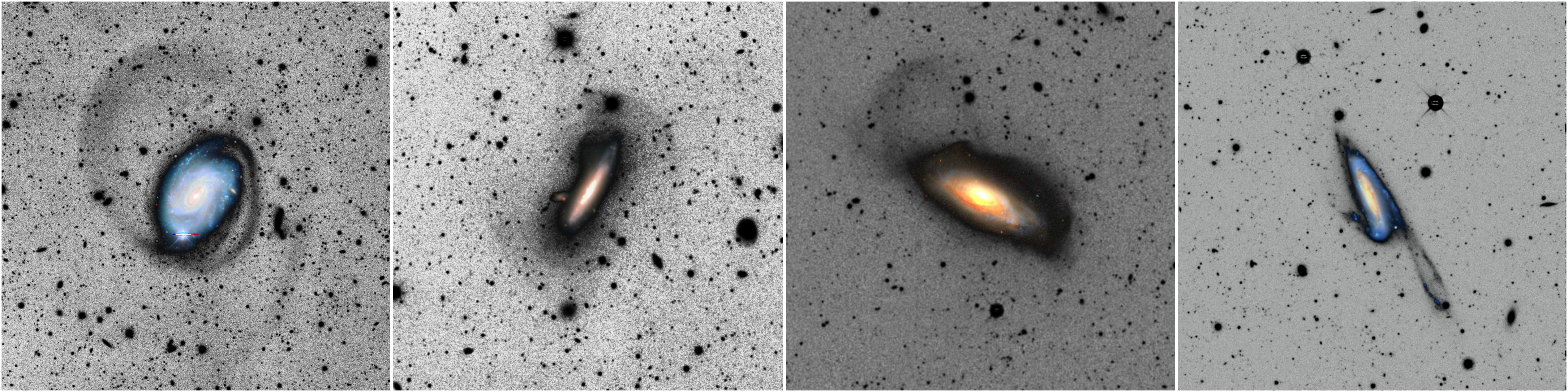}
	\caption{ Image cutouts of galactic feather candidates with diverse morphologies identified in the proof-of-concept phase of our survey (see Table~\ref{tab:target_list}): NGC 4981, NGC 5492, NGC 5635 and ESO245-010.}
    \label{fig-feathers}
\end{figure*}

\subsubsection{Galactic feathers}

Another possible source of confusion when classifying streams could come from mis-classification of perturbed stellar disc material in the form of tidal tails \citep{laporte19a} that can be excited through interactions with low-mass companions \citep{toomre72}. 

Minor mergers in late-type galaxies can excite various structures such as rings and tidal tails resembling structures seen at the disc-halo interface in the Milky Way \citep{kazantzidis09,slater14,gomez16,bauer18}. Among such structures showing stream-like morphology on the sky, we note the Eastern Banded Structure (EBS) and the Anticenter Stream (ACS) \citep{grillmair06}, which have long been suspected to be tidal debris of accreted satellites or massive globular clusters \citep[e.g.][]{penarrubia05, amorisco15}. However, recent $N$-body models of the interaction of a Sagittarius-like (Sgr) dwarf with the Milky Way have presented an alternative picture \citep{laporte18b}. In this model, structures like the ACS and EBS were re-interpreted as tidal tails in the outer disc resulting from the interaction with Sgr which, with later passages, were excited to larger heights, giving them the appearance of thin stream-like structures of disc material or ``feathers'' \citep{laporte19a}. A recent analysis of the chemical and age properties of the ACS supports this re-interpretation \citep{laporte20}. Due to the long orbital timescales in the outer disc ($t_{\mathrm{orb}}\geq1\,\rm{Gyr}$), these structures can survive and remain coherent for long timescales of the order $\sim6\,\rm{Gyr}$ or more and get gradually excited to larger heights above the Galactic midplane ($z\sim30\,\rm{kpc}$, making these potentially relevant sources of confusion as tidal streams in the stellar halo \citep{laporte19a}.

In some instances, ongoing mergers leave behind detectable signatures, such as an easily identifiable remnant core if the central galaxy is viewed at a favourable angle. However, in other merging configuration, it may be difficult to pinpoint whether a tidal stream could be a feather or the remains of a shredded satellite. 
%
Stars from the host can be distinguished by their ages and metallicities, inferred through their resolved stellar populations \citep[for example, from the ratio between M-giants and RR Lyraes,][]{price-whelan15}, although diagnostics that do not rely on spectroscopy or resolved star counts are very limited and have not been explored. In the case of the Milky Way, the ACS has a median metallicity $[Fe/H]\sim-0.7$ and magnesium abundance $[Mg/Fe]$, placing it at the $\alpha$-poor end of the distribution for thin disc stars with ages of $\tau>8\,\rm{Gyr}$ \citep{laporte20}, consistent with this structure originating from the disc. However, provided one has a constraint on the host's stellar halo mass, ($M_{\star}\sim10^{9}\,\rm{M_{\odot}}$ in the Milky Way \citep[see][]{deason19}, metallicity alone can suffice to rule out an ex-situ origin for substructures through the mass-metallicity relation of dwarf galaxies \citep{kirby13}. Thus, in the case of an external galaxy, provided colour information is available to derive robust photometric metallicities, it may thus be possible to distinguish streams from feathers using colour information alone in tandem with information on the total stellar halo mass of the system through deep exposures like those of this survey.

In addition to feathers, other diffuse overdensities may arise at the stellar halo-disc interface. An example is structures left behind by bending waves created during satellite encounters, which dissipate and flare the outer disc \citep{kazantzidis09, gomez13, chequers18, laporte18b}. Known cases in the Milky Way include the Monoceros Ring \citep{newberg02, slater14} or TriAnd \citep{rocha-pinto04}. Given their complex and amorphous shapes, such structures would be difficult to disentangle from minor merger events with photometry alone, and may ultimately require bespoke dynamical models of the outer discs of individual galaxies. Given the negligible self-gravity of the outer discs, it may be possible to use cost-effective modelling in the same spirit as Lagrangian Cloud Stripping methods \citep[e.g.][]{2014MNRAS.445.3788G} adapted to galactic discs.

In Fig.\,\ref{fig-cfpl}, we present a gallery of simulated feathers produced in Milky Way-like galaxies subjected to an ongoing merger with a 1:10 companion as presented in \citep{laporte18b}, viewed from different projection angles. It can be seen that, depending on the viewing angle, feathers can exhibit a wide range of complex morphologies which may lead to confusion with the features generated by dwarf galaxy disruption. In Fig.~\ref{fig-feathers}, we showcase possible candidates of `in situ´ feathers in our proof-of-concept sample. 

Although it remains to be seen if colour information alone is sufficient to reliable distinguish feathers from streams, our survey provides a useful test bed to systematically study in-situ structures at the disc-halo interface in external galaxies, and an opportunity to devise tests to disentangle them from genuine streams of accreted substructures (e.g. through residuals from direct image modelling).

\subsection{Disentangling tidal features from Galactic cirri}
\label{sec:cirrus}

In this section we describe a potentially useful test to identify false detections due to Galactic dust, which often mimics the morphology and surface brightness of tidal features in deep images. Our selection criteria reject galaxies in the zone of avoidance (Galactic latitudes $|b| < 20\deg$). However, low-surface brightness, optically-thin dust clouds at relatively high Galactic latitudes are expected to be detected in our deep data.

Since both cirrus and diffuse features have a wide range of properties, it is helpful to use multiple diagnostics to distinguish them. First of all, visual analysis of the images is often sufficient. Most tidal feature morphologies are unambiguous and, together with the absence of structures of similar surface brightness in the field surrounding the host galaxy, allow us to rule out false positives. Also, detections of dust features in far infrared (FIR) data, such as those from {\it Herschel} \citep[or in its absence, IRAS 100 $\mu$m or Planck 857 Ghz,][]{miville_2005} can be used.
Recent work by \citet{2020A&A...644A..42R} provides another diagnostic to distinguish tidal streams from Galactic cirrus clouds, which makes use of the optical colours of the cirri in the $g$, $r$, $i$ and $z$ bands. \citet{2020A&A...644A..42R} show that the colours of the cirrus clouds are significantly different from those of any galactic stellar population, being characterised by a bluer $r-i$ colour for a given $g-r$ colour. This method has the great advantage of being based on the optical observations themselves, independent of complementary IR data which may not be available in all cases. This diagnostic also benefits from the high spatial resolution in our optical images. 

\begin{figure}
	\includegraphics[width=1.0\columnwidth]{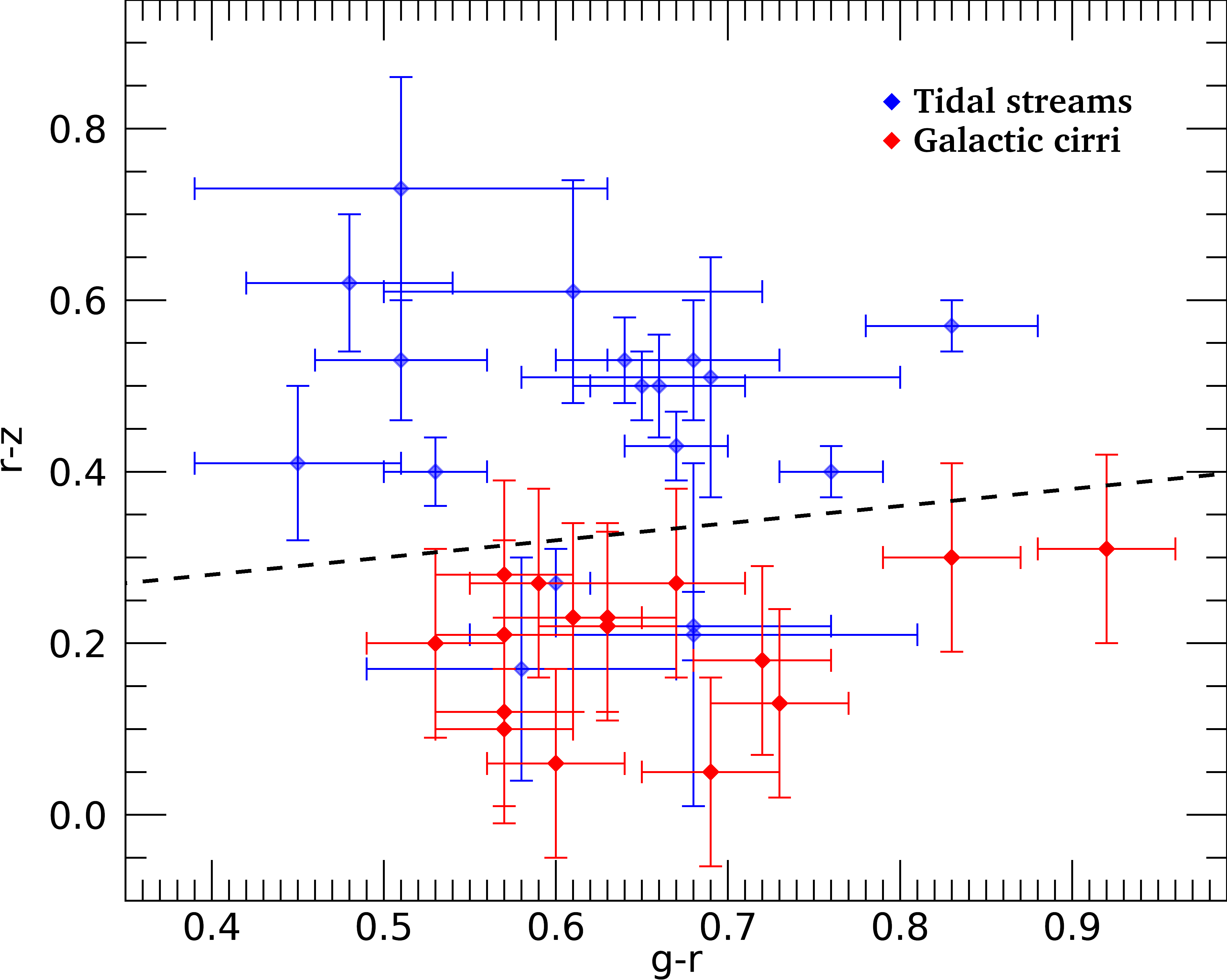}
    \caption{g-r vs. r-z colours for the tidal streams presented in Table \ref{tab:streams_photo} (we exclude tidal features with colour errors > 0.25 mag for clarity) and the Galactic cirri regions characterised by \cite{2020A&A...644A..42R}. The black dashed line, $(r-z) = 0.20 \times (g-r) +0.20$, limits the regions where the colours are compatible/incompatible (below/above of the line) with Galactic dust.}
    \label{fig-tidals_dust}
\end{figure}

In Fig.~\ref{fig-tidals_dust}, we plot the colours of the features described in Table \ref{tab:streams_photo} against the colours of Galactic cirri reported in \citet{2020A&A...644A..42R}. The figure shows that the colours of cirri are mostly clustered in a relatively compact region of $g-r$ vs. $r-z$ space. In contrast, we see that the colours of the tidal streams on this map occupy a much larger area. Although the two colour distributions partly overlap (relative to \citet{2020A&A...644A..42R}, the $grz$  bands are not optimal for this separation), the separation is nevertheless significant enough to provide a useful estimate of the probability that a given feature is due to cirrus.

Colours  $(r-z) > 0.20 \times (g-r) + 0.20$ (see Fig. \ref{fig-tidals_dust}) are highly likely to indicate diffuse extragalactic stellar populations rather than Galactic dust. We caution that, although potentially useful, this method is by no means definitive. It can be applied as a last resort in the absence of additional data, but it is most powerful in combination with other more complex approaches.

\begin{table*}
\centering
\caption{Comparison between the Legacy Survey and future surveys.}
\label{tab:comparison}
\begin{tabular}{lcccc}
\hline
Survey & Area (deg$^{2}$) & S.B.$^{1}$ & Pixel scale ($''$) & Band(s) ($\mu$m)\\
\hline
DESI LS & 14,756 & 28.5$^{2}$  & 0.26 &  0.5 - 1.0 \\
Rubin LSST & 18,000 & 29.55$^{3}$ & 0.2 & 0.3 -- 1.1 \\
Roman (NIR) & $>$\,1700 & 30.5$^{4}$ & 0.11 & 0.9 -- 2.0 \\
Euclid (VIS) & 15,000 & 29.5$^{5}$ & 0.1  & 0.5 -- 0.9 \\
Euclid (NIR) & 15,000 & 28.4$^{6}$ & 0.3  & 0.9 -- 2.0 \\
Euclid Deep (VIS) & 43 & 31.5$^{7}$ & 0.1 & 0.5 -- 0.9 \\
\hline
\multicolumn{5}{l}{$^{1}$Limit AB mag arcsec$^{-2}$ in 10$''$ $\times$ 10$''$ boxes, 3$\sigma$.} \\
\multicolumn{5}{l}{$^{2}$Estimate in $r$-band from this paper.} \\
\multicolumn{5}{l}{$^{3}$Estimated for Rubin final mosaics after 10-year integration} \\
\multicolumn{5}{l}{$^{4}$Estimate based on a 30-minute long total integration time except in the} \\ \multicolumn{5}{l}{F213 filter where the estimate is 28.8 due to thermal contamination.} \\
\multicolumn{5}{l}{$^{5}$Euclid VIS estimates are from \citet{euclid22a}.} \\
\multicolumn{5}{l}{$^{6}$Euclid NIR estimates are from \citet{euclid22b}.} \\
\multicolumn{5}{l}{$^{7}$On average, varying between 31 and 32.}
\end{tabular}
\end{table*}


\subsection{Comparison with existing and future ultra-deep imaging of nearby galaxies}
\label{sec:othersurveys}

The analysis of low surface brightness observations is hindered by the lack of consensus on the most appropriate method for
measuring the surface brightness limit of an astronomical image \citep{mihos_2019}. One of the most common methods is to measure the width of the background noise distribution over pixels of a fixed size \citep[e.g.][]{{trujillo_2016}}. We follow this approach throughout, by quoting surface brightness limits (i.e. residual sky backgrounds, depths of our mosaics) as 3$\sigma$ of the surface brightness distribution measured over an area of $10\arcsec\times10\arcsec$. This method only takes into account random statistical noise, not systematic errors. Images suffering from severe over-subtraction of the sky background have the same surface brightness limit as well-calibrated images according to this definition \citep{borlaff_2019}.



\begin{figure}
 \begin{center}
 \includegraphics[trim={12 12 12 52}, clip, width=0.5\textwidth]{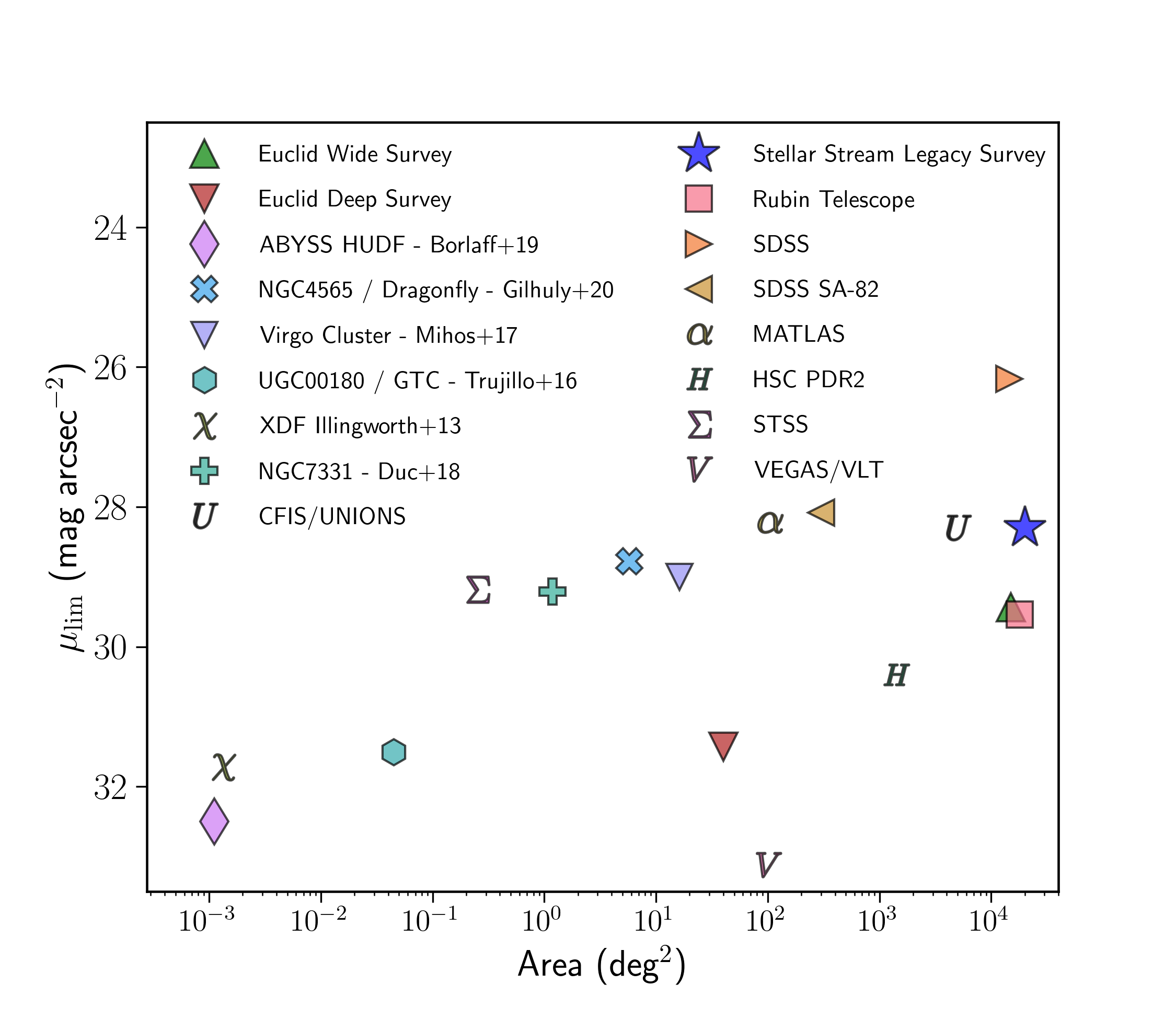}
\caption{Comparison of the surface brightness limit ($3\sigma$, $10\times10$ arcsec$^2$) as a function of the observed area for a selection of deep optical and NIR surveys. Depths are measured in the $r$-band (noted when not applicable). The {\it Stellar Stream Legacy Survey} presented in this paper is indicated with a blue star. The other surveys comprise: Stellar Tidal Stream Survey \citep[STSS]{martinezdelgado_2019}, \textit{Euclid} VIS \citep{euclid22a}, SDSS \citep{york_2000}, IAC Stripe 82 \citep[][]{IACStripe82}, the MATLAS deep imaging Survey \citep[$g$-band magnitudes][]{duc_2015}, Hyper Suprime-Cam DR2 \citep{Aihara2018}, NGC 4565 Dragonfly observation \citep{gilhuly_2020}, HST WFC3 ABYSS HUDF \citep[NIR bands][]{borlaff_2019}, XDF \citep{illingworth_2013}, UGC00180 10.4m GTC exploration \citep{trujillo_2016}, the Burrell Schmidt Deep Virgo Survey \citep{mihos_2017}, the CFIS/UNIONS survey \citep{Sola2022}, the VEGAS/VLT survey \citep{Ragusa2021}, and the \textit{Rubin Telescope} Legacy Survey of Space and Time \citep[10 years full survey integration,][]{ivezic_2019}.}
\label{fig:depth_vs_area}
\end{center}
\end{figure}

\begin{figure*}
	\includegraphics[width=1.0\textwidth]{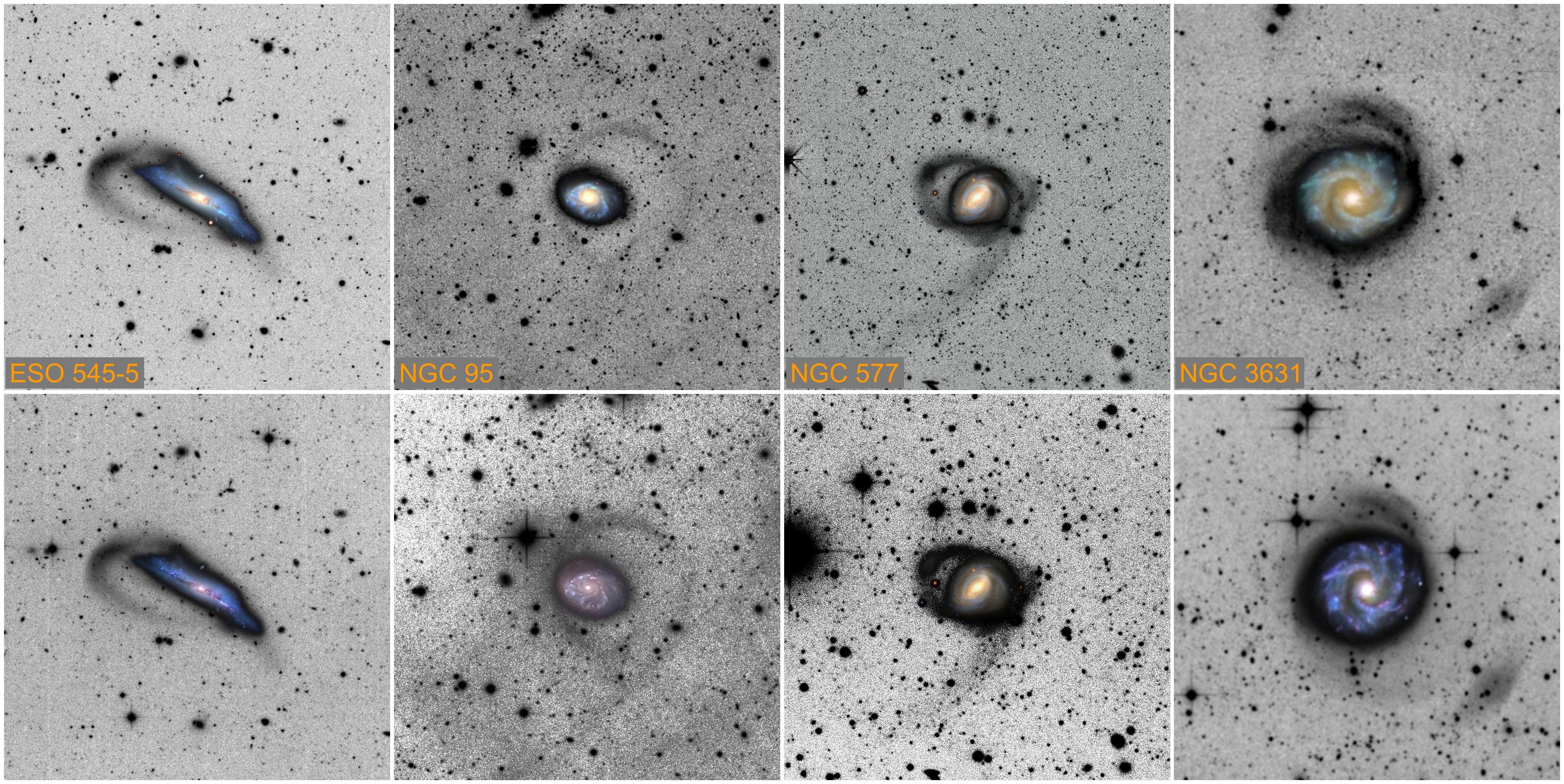}
          \caption{Comparison of the image cutouts obtained by applying the methodology described in this paper to data from the {\it DESI Legacy Imaging Surveys} ({\it top panels}) versus long exposure times (6--10 hours) images obtained with amateur robotic telescopes from the Stellar Tidal Stream Survey \citep[][{\it bottom panels}]{martinezdelgado_2019}: our new {\it Stellar Stream Legacy Survey} is well positioned to uncover hundreds of low surface brightness features in the local Universe.}
   \label{fig-amateur}
\end{figure*}

In Fig.\,\ref{fig:depth_vs_area} we compare the surface brightness magnitude limit (calculated as above, and averaged over all the available filters for each field or survey) as a function of the observed area for several current and planned deep imaging surveys. Extremely deep surveys like the Hubble Ultra Deep Field \citep{illingworth_2013,borlaff_2019} have a much smaller area (a few arcmin$^{2}$) compared to extremely wide but shallower surveys like SDSS \citep{york_2000}. For the time being, depths below $30$ \magarc\ are incredibly hard to achieve with ground-based telescopes \citep{trujillo_2016} and prohibitive for a large number of galaxies. Fig.~\ref{fig-amateur} shows that there are no striking differences in limiting depth between images taken with amateur telescopes from the Stellar Tidal Stream Survey \citep{martinezdelgado_2019} and the results of our custom pipeline for the new {\it Stellar Stream Legacy Survey} applied to data from the {\it DESI Legacy Imaging Surveys}. Ten minutes of computing time to reprocess images from the latter can achieve the same depth as $\sim$10 hours of dark observing time with amateur telescopes for each galaxy. A survey using small telescopes equivalent to compiling a sample of $\sim$3000 galaxies out to 100 Mpc would need $\sim$20\,000 hours of dark time. These results support our strategy of using archival {\it DESI Legacy Imaging Surveys} data, which also have superior seeing. 


A number of other deep imaging surveys are expected to approach $30\,\mathrm{mag\,arcsec^{-2}}$ in the next decade, including the space-based 1.2-m visual to NIR {\it Euclid} space telescope, the ground-based visual wavelength 8.4-m Vera C. Rubin Observatory \citep[][hereafter Rubin]{ivezic_2019} and the space-based 2.4-m, NIR {\it Nancy Grace Roman Space Telescope}. The expected sensitivities and other basic parameters are compared to the Legacy Survey in Table \ref{tab:comparison}. Points corresponding to the depths of the \textit{Rubin} and \textit{Euclid} surveys (according to our standard definition) appear in the low right corner of Fig.~\ref{fig:depth_vs_area}.

Although these future surveys do not cover a significantly larger area than our survey, they will eventually provide a significant improvement in depth and spatial resolution. Greater depth allows the detection of fainter stellar streams arising from ultra-faint satellites and globular clusters \citep[see, e.g.][]{Grand17}. Higher spatial resolution permits more accurate masking of distant background galaxies, and hence aids the recovery of thin stellar streams from globular clusters \citep[see][]{pearson_2019}. Improved masking of superposed sources could also increase the accuracy of colour (hence age and metallicity) measurements for streams. Deep NIR images improve stellar mass estimates and population constraints for streams, by sampling the peak of their SEDs. 
\section{Outlook and comparison with cosmological simulations}

\label{sec:discussion}

 In the previous sections we have described the ideas and techniques of our new {\it Stellar Stream Legacy Survey} and we have shown proof-of-concept results for 24 new stellar streams detected in the local Universe.  In this section, we discuss how the depth, coverage and uniformity of the full survey dataset can be exploited to address two key methodological and scientific problems related to stellar streams that we plan to undertake in future: automatic detection of stellar streams, morphological classification and comparison to cosmological simulations. 

\begin{figure*}
\includegraphics[width=1.0\textwidth]{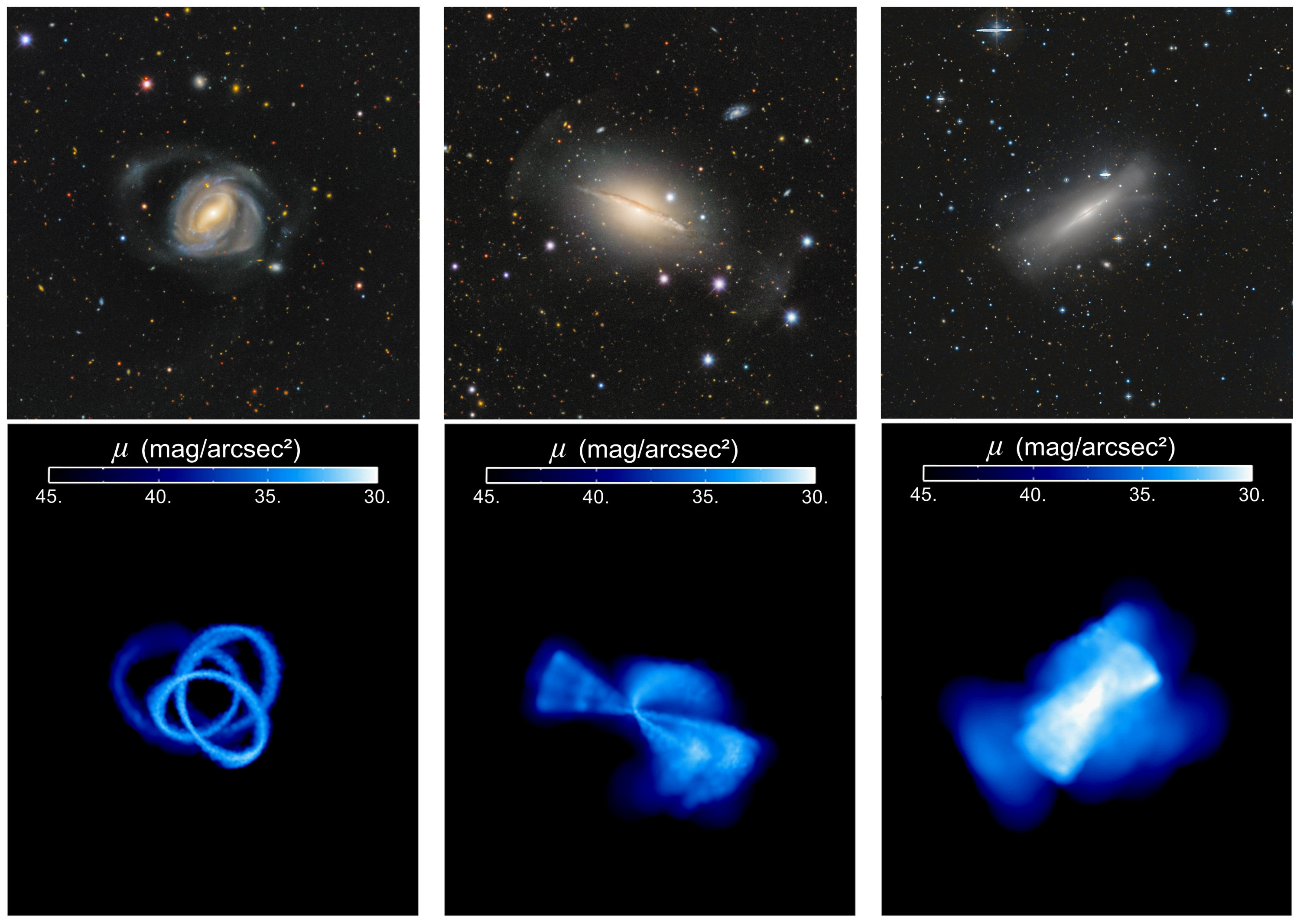}
   \caption{ {\it Bottom panels}. External perspective snapshots for the three morphological types classified by \citet{johnston_etal_2008}: “Great circle” streams ({\it left}) originated from satellites accreted 6--10 Gyr ago on nearly circular orbits; “shells” or “umbrellas” ({\it middle}) arise from accretion events within the past $\sim$8 Gyr with radial orbits; and “mixed”-type tidal remnants ({\it right}) formed from ancient (more than 10 Gyr ago) accretion events that have fully mix along their orbits. {\it Top panels}: image cutouts obtained from the {\it DESI Legacy Imaging Surveys} data for each morphological case of tidal stream: a great circle stream in NGC~577 ({\it left}), shell-like features around NGC~681 ({\it middle}); and a “mixed” remnant around NGC~5866 ({\it right}). Figure adapted from \citet[][]{carlin_2016}.}
\label{fig-bjstreams}
\end{figure*}

\begin{figure}
	\includegraphics[width=1.0\columnwidth]{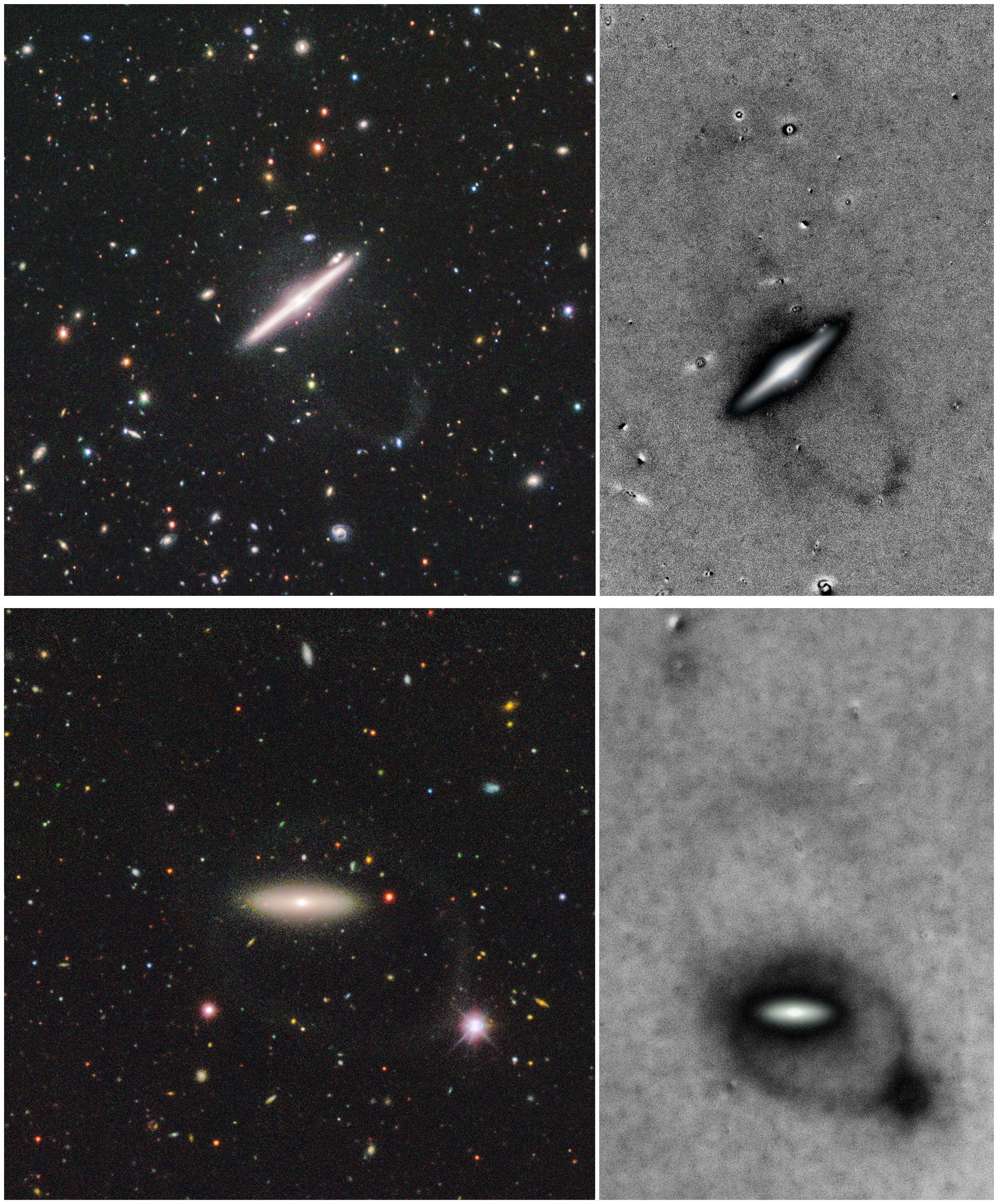}
        
    \caption{Evidence of very faint, extended outer loops around the edge-on spirals UGC08717 ({\it top panels}) and 2MASXJ12284541-0838329 ({\it bottom panels}) detected as a result of the accurate sky background subtraction described in Sec.~\ref{sec:cutout_method}. These examples illustrate the effect of the surface brightness cut-off of the data on the interpretation of the overall morphology of the streams.}
    \label{fig-outerloops}
\end{figure}

\subsection{Automatic detection of stellar streams with Deep Learning algorithms}

We complement our visual identification of tidal streams by employing convolutional neural networks \citep[CNN;][]{2014arXiv1409.1556S} to detect stellar tidal streams. CNNs, a form of deep learning \citep[e.g.][]{2015Natur.521..436L}, take multi-dimensional data as an input, such as an image, and perform a series of non-linear convolutions before outputting a classification. This technique has been increasingly used in astronomy in recent years to identify rare objects in imaging surveys, such as quasars \citep[e.g.][]{2018A&A...611A..97P}, strong lenses \citep[e.g.][]{2017MNRAS.472.1129P, 2019MNRAS.487.5263D} and galaxy mergers \citep[e.g.][]{2018MNRAS.479..415A, 2019A&A...631A..51P}. Classification of mergers, in particular, relies on the identification of diffuse features \citep{2019A&A...626A..49P}, demonstrating that this technique is ideally suited to identifying stellar streams. CNNs also have the advantage of being reproducible: the same classification (stellar stream or no stellar stream, in this case) is reported every time the image is passed through the network.

\citet{walmsley_2019} presented the first attempt to identify faint tidal features around galaxies from the CFHTLS-Wide Survey using CNNs trained on a relatively small number of streams (305).
The models achieve 76\% completeness and 20\% contamination. This relatively poor performance, compared to other classification tasks for galaxy images obtained with CNNs, may be due to the combination of the faintness of the tidal features combined with the lack of a large training sample.

To address the limitation imposed by the small size of the sample of known tidal streams, our survey trains the CNN using mock images generated from particle-spray models (see Sec.~\ref{sec:morphological}). Accurate classification of galaxy mergers with CNNs has been demonstrated using a few thousand merger examples \citep[e.g.][]{2018MNRAS.479..415A, 2019A&A...631A..51P,2021MNRAS.504..372B}. As our objective is arguably more challenging than merger classification, we generate of the order of ten thousand mock tidal streams. These mock streams are combined with an identically sized sample of mock galaxies without streams, to provide a class-balanced training sample. As tidal streams are rare, a sample of galaxies that is matched to the true ratio of galaxies with and without tidal streams runs the risk that the CNN never learn to identify the features of interest.  Performance metrics can be improved by assigning all objects to the majority class.

The mock images used for training are single channel to prevent any incorrect or non-physical colour information from the particle-spray model influencing the detection of streams, choosing the deepest band within each survey to maximise our sample size. Alternatively, it is possible to normalise each band independently, as done in \citet[][]{2018MNRAS.476.3661D} and \citet[][]{vega-ferrero_2021}, which again remove any colour dependence of the classification. Once the CNNs are trained, they are able to identify the stellar streams in a fraction of the time it would take a human. Moreover,  as demonstrated in \citet[][]{vega-ferrero_2021}, machines are able to recover features hidden to the human eye when trained on a sample with correct labels (e.g. with labels coming from simulations). This technique allows the detection of faint tidal stream structures which may otherwise be missed by visual inspection.

Due to the differing angular resolutions, we train one CNN per survey. Training multiple CNNs has the benefit that it is possible to generate a sub-sample of very robust stellar streams in the regions where the surveys overlap. This is done by selecting galaxies that are identified as having stellar streams by the CNNs for two, or more, different surveys. 

The performance of deep learning models trained with simulations and applied to real galaxies can be affected if the simulations do not capture the full complexity of the real images (e.g. noise, PSF, seeing and other observational effects).The mock images reproduce the processing steps applied to the real images as closely as possible. This includes applying a realistic noise model and convolving with the PSF of the survey in each band. Such techniques have also been demonstrated in the context of galaxy merger detection \citep{2019A&A...626A..49P, 2020A&A...644A..87W, 2021MNRAS.504..372B}. Domain adaption is also be used, a technique in which a CNN is paired with a Generative Adversarial Network to force the CNN to give identical, or at least similar, results for the simulation and real images \citep[e.g.][]{WANG2018135}. Combining simulations and real images in this way help ensure the reliability of results obtained by applying the CNN to the real survey data. The sample of tidal streams identified by the {\it Stellar Stream Legacy Survey} is, in turn, useful for training of future CNNs for the mophological analysis of galaxies.

\subsection{Morphological classification of stellar streams}
\label{sec:morphological}

\citet[][]{johnston_etal_2008} showed that estimates of the frequency, sky coverage and fraction of stars residing in substructure in stellar halos can be used as a diagnostic of recent merger events. In particular, they found that the surface brightness and morphology of substructure can constrain the masses and orbital distributions of recently accreted satellites. \cite{2015MNRAS.454.2472H} found that the observed ratio between the number of {\it great circle streams} (formed by disrupted satellites on near-circular orbits) and the number of {\it umbrellas/shells} (formed by satellites on more radial orbits) can distinguish between the infall distributions predicted by different cosmological simulations, and that meaningful constraints could be obtained by an unbiased, uniform survey containing a few hundred stellar streams. Our survey yields even better statistics for substructure morphologies in the local Universe and hence place tighter constraints on the predictions of $\Lambda$CDM simulations. Fig.~\ref{fig-bjstreams} shows examples from our {\it Stellar Stream Legacy Survey} images compared to the theoretical morphological classification of tidal debris by \citet[][]{johnston_etal_2008}.




However, we first need to characterise hundreds of debris structures in our survey. 
\citet[][]{hendel_2019} developed a new machine vision technique which can automatically differentiate between shell-like and stream-like features through identification of density ‘ridges’ that define substructure morphology, which we can apply to our image data. More generally, 
our survey is well placed to make progress on techniques for the automatic morphological classification of streams. We therefore use two different approaches to characterise debris structure and streams in our sample:\\

\noindent {\bf Visual classification}: Diffuse structures in stellar halos are typically classified into three morphological types: {\it streams}, {\it umbrella/shells} and {\it mixed} \citep[see e.g.][]{johnston_etal_2008}. There are currently few methods that can surpass visual inspection for the identification of spurious detections. It is important to note that surveys are often not sensitive to the most diffuse parts of the debris structure. The two panels of Fig.~\ref{fig-outerloops} illustrate this by showing additional structure revealed after an accurate sky background subtraction. Additionally, streams on specific orbits can lead to `fans' of low-density tidal debris \citep[][]{pearson_2015, 2016MNRAS.455.1079P, 2021MNRAS.501.1791Y}, which are potentially missed. In our survey, we therefore use visual classification to exclude
amorphous, diffuse structures in our images that are likely to correspond to more ancient accretion events or map out complex orbits in their host galaxy.

\vspace{0.1in}

\looseness=-1
\noindent {\bf `Particle spray' library and fits:} Bespoke $N$-body simulations can be used to generate streams that provide a qualitative resemblance to observed features \citep{martinezdelgado_2008}. However, a full exploration of a wide parameter space using $N$-body simulations is usually too expensive. In our survey, we instead use a `particle spray' technique to rapidly generate an extensive library of mock streams \citep[see e.g. ][]{bonaca_2014,kuepper_etal_2015,2014MNRAS.445.3788G,fardal2015}. In the `particle spray' technique, the stars, which escape a progenitor to form leading and trailing stellar arms, are massless test particles. Test particles are released uniformly in time from the progenitor's Lagrange points, whose locations are calculated based on the progenitor's initial mass. The progenitor mass can be updated throughout the simulation to mimic its mass-loss. 
Once the test particles are released from the Lagrange points with a dispersion in the velocity with respect to the progenitor, the positions and velocities of the test particles are evolved in the combined gravitational potential of the host and the progenitor. \cite{2014MNRAS.445.3788G} showed that including the self-gravity of the progenitor is essential for the `particle spray' technique to reproduce the length of streams generated with $N$-body simulations. The fact that the test particles do not mutually interact with each other makes this technique extremely fast compared to direct $N$-body simulations. The morphology the resulting mock streams compare remarkably well to those generated with direct $N$-body simulations (see e.g. \citealt{2014MNRAS.445.3788G}). 

We plan to use the modified Lagrange Cloud stripping variation of the `particle spray' technique presented in \cite{2014MNRAS.445.3788G}, an example of which is shown in Fig.~\ref{fig:stream_spray}  (right),
to generate a large library of mock tidal disruptions with orbits and masses based on distributions obtained from the cosmological simulations described below. 
`Particle spray' simulations are carried out in a range of galactic potentials which span the types of galaxies in our survey (i.e. various mass profiles and dark matter halo shapes). Mock debris structures in this library are compared visually to our observations of tidal debris. In particular, we can inject mock debris into real images and test how well our visual inspection works. For example, if a shell-like feature is sufficiently faint, it is possible that only the rim of the shell would be visible above the background (e.g. the bottom, middle panel of Fig. \ref{fig:Fig3A_July31}). These tests indicate whether such features can be distinguished from a thinner tidal stream on a near-circular orbit. 

In addition to constraining the build-up of stellar halos through comparisons to cosmological models, we identify a subset of cases in our survey with prominent streams and use these to fit the potential of individual host galaxies. While many Galactic streams have been used to constrain the potential of the Milky Way \citep[e.g.][]{2010ApJ...712..260K,2010ApJ...714..229L,kuepper_etal_2015,bovy_et_al_2016,malhan_etal_2019}, this has only been attempted for two external galaxies \citep[e.g.][]{2013MNRAS.434.2779F, 2015arXiv150403697A}. The `particle spray' technique can also be used to determine which mass distributions best reproduce the properties of observed streams in the Legacy Survey \citep{2014MNRAS.445.3788G}. This technique has  already been applied to map the mass distribution in the Milky Way  \citep[e.g.][]{2019MNRAS.487.2685E}. Constraints on the dark matter potentials of survey galaxies from this technique can be contrasted with those obtained through satellite kinematics and used to constrain relationships between galaxy properties and halo mass, which are fundamental to the galaxy formation models described in the following section.


\subsection{Comparison with cosmological simulations}
\label{sec:cosmosim}



 
 
 \begin{figure*}
	\includegraphics[width=0.3\linewidth]{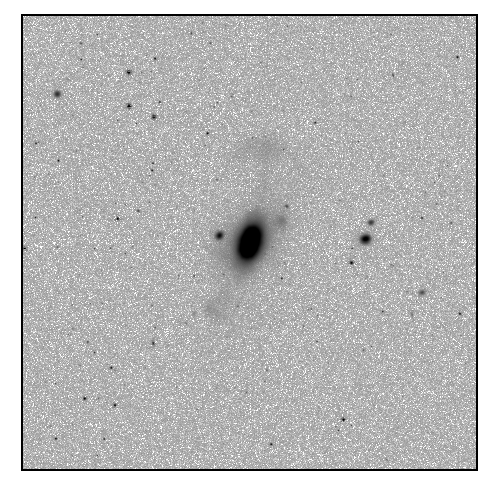}
	\includegraphics[width=0.3\linewidth]{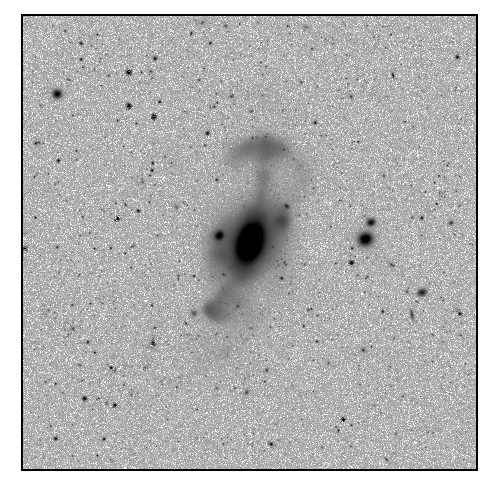}
	\includegraphics[width=0.3\linewidth]{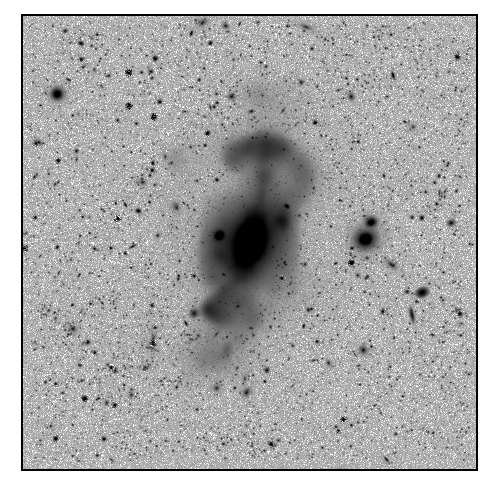}

	\includegraphics[width=0.3\linewidth]{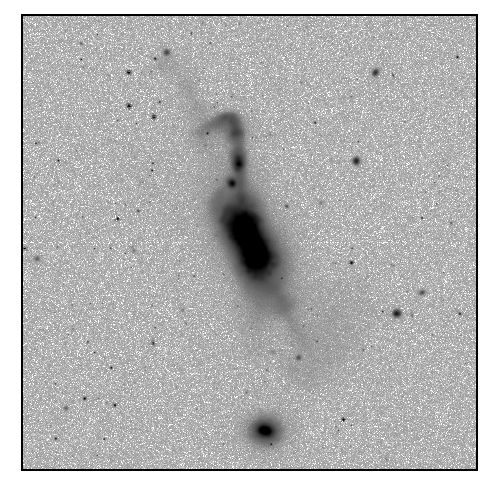}
	\includegraphics[width=0.3\linewidth]{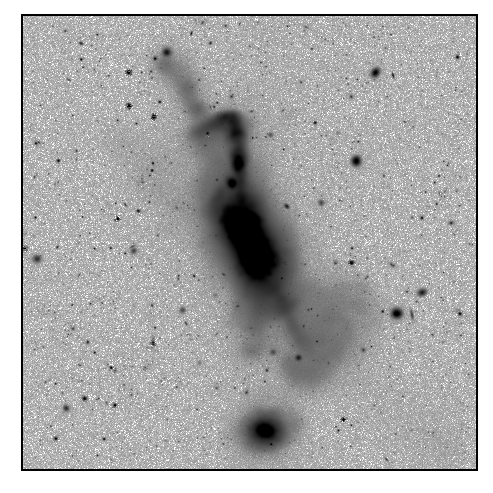}
	\includegraphics[width=0.3\linewidth]{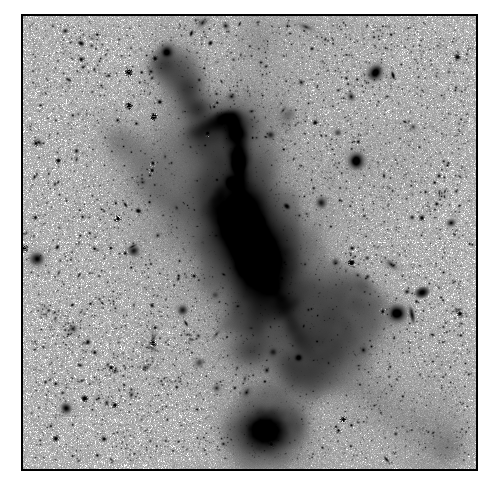}
		
	\includegraphics[width=0.3\linewidth]{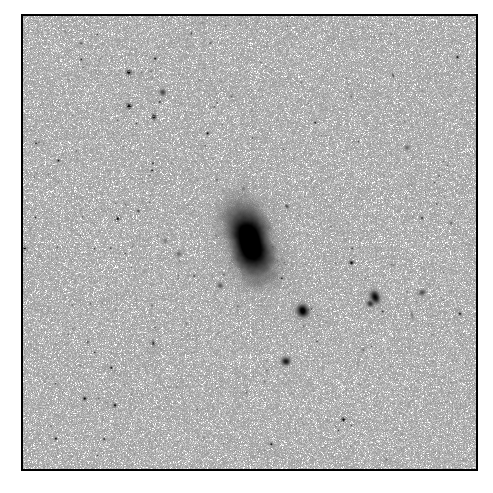}
	\includegraphics[width=0.3\linewidth]{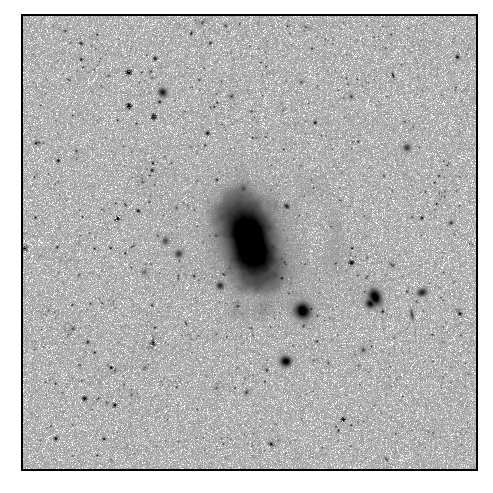}
	\includegraphics[width=0.3\linewidth]{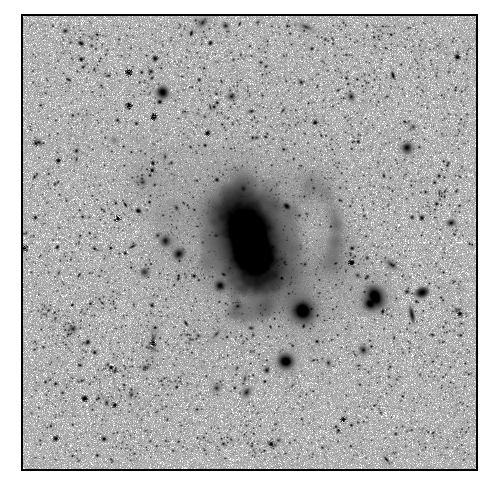}

    \caption{Example $r$-band mock images of three $\sim L_{\star}$ galaxies from \coco{} at a fiducial distance of 50~Mpc From left to right, the images have surface brightness limits of $\mu_{\mathrm{lim}}=27$, $29$ and $31$ magnitudes per square arcsecond, respectively. The dynamic range of each image runs from $\mu_{\mathrm{lim}} + 1$ (white) to $\mu_{\mathrm{lim}}-9$ (black).  We have embedded these models in the same fiducial field from the {\it DESI Legacy Imaging surveys}.
Further details of how we generate these mock images are given in the text. From top to bottom, the stellar masses of the galaxies are  $\log_{10} M_{\star}/\mathrm{M_{\odot}} = (10.1, 10.6, 10.1)$, and their virial masses are $\log_{10} M_{\mathrm{vir}}/\mathrm{M_{\odot}} = (11.8, 12.2, 12.2)$.}
    
    \label{fig:coco_example}
\end{figure*}

In the previous sections we have demonstrated that all-sky imaging surveys now reach the depth required to assemble statistically representative samples of diffuse tidal features in the Local Universe, and to quantify their frequency of occurrence, mass, morphology, spatial scale and stellar populations. Cosmological simulations are the only practical way to interpret the observed distributions of these properties. In turn, such observations can constrain the `subgrid' models on which the simulations are built.

To exploit the statistical power of our survey, cosmological simulations suitable for comparison must have comparable volume and dynamic range from which to draw representative samples of galaxies. The physical models underlying these simulations are typically calibrated against statistical observables of the galaxy population, such as the stellar mass function, which allows the same selection function to be applied to both observed and simulated galaxies.

For a thorough comparison with our survey data that takes detectability and projection effects into account, it is necessary to derive realistic mock images \citep[e.g.][]{torrey_2015,snyder_2015} from a range of different simulations. These are stringent requirements, but nevertheless well-matched to the capabilities of state-of-the-art simulations from multiple groups. Below we describe the potential for comparison between our {\it Stellar Stream Legacy Survey} data and two state-of-the-art cosmological simulations, the CoCo and TNG50 simulations, and illustrate the techniques for generating mock images.

Cosmological simulations can be used to quantify intrinsic galaxy-to-galaxy scatter in the properties of streams, for comparison to observed distributions. In addition, we seek to quantify  variance in the predicted structure of stellar streams and the disruption times of satellites between different galaxy formation models and simulation codes. For this purpose, we use the cosmological runs of the AGORA project\footnote{\url{https://sites.google.com/site/santacruzcomparisonproject/}}. AGORA is a  comparison project exploring code-to-code (or group-to-group) differences in galaxy simulations \citep{Kim2014, Kim2016}. In its third phase, the AGORA community is comparing zoom-in cosmological simulations of the same $M_{\mathrm{vir}}\sim$10$^{12}\,\mathrm{M}_{\sun}$ halo at redshift zero, with gravitational softening length $\epsilon=80$~pc, and particle mass resolutions of $M_\mathrm{DM}=2.8\times 10^{5} \, \mathrm{M_{\odot}}$ and $M_{\star|\mathrm{gas}}=5.65\times 10^4 \,\mathrm{M_{\odot}}$ (Roca-F\`abrega et. al in preparation). Using these simulations, we explore the numerical and resolution dependence of predictions for observables relevant to our tidal stream survey, with the ultimate goal of finding new ways in which the survey data can be used to constrain subgrid models.

\subsubsection{The Copernicus Complexio models}

The \textit{Copernicus Complexio} (\coco{}) cosmological $N$-body simulations \citep{hellwing_2016} provides both high mass resolution ($1.6\times10^{5}\mathrm{M_{\odot}}$ per particle) and a representative analogue of the Local Volume (a high-resolution spherical region of radius $\sim25$~Mpc embedded in a lower-resolution box of side length $100$~Mpc), well suited to comparison with the {\it Stellar Stream Legacy Survey}.

We have applied the technique known as particle tagging to \coco{}, following the \texttt{STINGS} methodology of \citet{cooper_2017}, building on previous applications to the Aquarius and Millennium II simulations \citep{cooper_2010,cooper_2013}. By coupling collisionless simulations to semi-analytic models of galaxy formation, \texttt{STINGS} creates dynamically self-consistent cosmological $N$-body simulations of galactic stellar halos and their associated tidal streams, at a resolution comparable to current hydrodynamical simulations and without many of the simplifications required by a earlier generation of `fast but approximate' galaxy models \citep[e.g.][]{bullock_johnston_2005}. In this example we have used the \texttt{Galform} semi-analytic model \citep{lacey_2016} to predict the evolution of stellar mass, size, and chemical abundances in every \coco{} dark matter halo, constrained by comparisons to the large-scale statistics of the cosmological 
galaxy population (for example, optical and infrared luminosity functions).  Individual dark matter particles in \coco{} were then ``painted'' with single-age stellar populations according to a dynamical prescription that specifies the binding energy distribution of stars at the time of their formation (this prescription differs between particle tagging implementations by different groups; the details for \texttt{STINGS} and a comparison to other approaches are given in \citealt{cooper_2017}). Using the tagged particles, the individual star formation history of each satellite (and hence properties such as stellar mass, luminosity and metallicity) can be studied alongside the full phase-space evolution of its stars, before and after tidal disruption. Particle tagging can be thought of as an extension to the semi-analytic models that allows individual stellar populations to be tracked in phase space, while retaining the computational efficiency and flexibility of the semi-analytic galaxy formation model.

These simulations have sufficient dynamic range both to draw representative samples of galaxies and to study debris features from dwarf galaxies comparable to the classical satellites of the Milky Way. Importantly, \coco{} adequately resolves the cores of massive satellite halos in these systems and the fine structure of the debris from their tidal disruption. Their
softening scale of $330$ pc (comparable to the width of the Milky Way's
Orphan stream) and particle mass resolution of $1.6\times10^{5}
M_{\odot}$ are similar to the
Medium-Resolution suite of the Apostle Milky Way Zoom simulations
(comprising 11 galaxies). 
\coco{}, however, provides a representative cosmological
volume containing $\sim$50--70 Milky Way mass dark matter halos. Assuming a total-mass-to-light ratio of 100 in the region of
phase space corresponding to the stellar body of a dwarf like the
Fornax dSph, its debris would be resolved with several thousand tagged
particles in these simulations.

Fig.~\,\ref{fig:coco_example} shows mock images of three approximately $L^{\star}$ galaxies from \coco{}, embedded in a simulated field constructed using {\it DESI Legacy Imaging Surveys} model fits to all sources in a random patch of sky. These examples adopt a fiducial distance of 50 Mpc. From left to right in Fig.~\ref{fig:coco_example}, we vary the surface brightness limit to simulate, respectively, an SDSS-like observation ($\approx$ 27 \magarc), a {\it Stellar Stream Legacy Survey} observation ($\approx$ 29 \magarc) and an even deeper image comparable to what may be achievable with HSC or \textit{Rubin} ($\approx$ 31 \magarc). In the first two rows, new features appear in the {\it Stellar Stream Legacy Survey}-like images that are not visible at the SDSS depth, and these are further enhanced (hence easier to interpret) at HSC-like depth.



To create these images, we first smooth the projected $N$-body tagged particle distribution in a box of side length $500$~kpc around each galaxy, using a conventional two-dimensional `SPH' cubic spline kernel. The smoothing length of each tagged particle is proportional to the RMS distance to its 16$^\mathrm{th}$ nearest neighbour\footnote{Each stellar population associated with a single $N$-body particle is smoothed by the same amount. This is for cosmetic purposes and the optimal choice of smoothing length is somewhat arbitrary; \citep[see e.g.][]{lowing_2011} for a more rigorous approach to smoothing in mock observations of a sparsely sampled stellar phase space.}. This smoothed image is further convolved with a simple analytic approximation to the Legacy Surveys PSF. Gaussian noise is added to simulate a sky-subtracted image of the required depth. We use the same definition of limiting surface brightness, \mulim, as elsewhere in this paper, using $3\sigma$ of the flux distribution in $10\arcsec \times 10\arcsec$ regions. Finally, we superpose the Tractor models of all sources from the {\it DESI Legacy Imaging Surveys} catalogue in a random patch of blank sky.


\begin{figure*}
	\includegraphics[width=18cm]{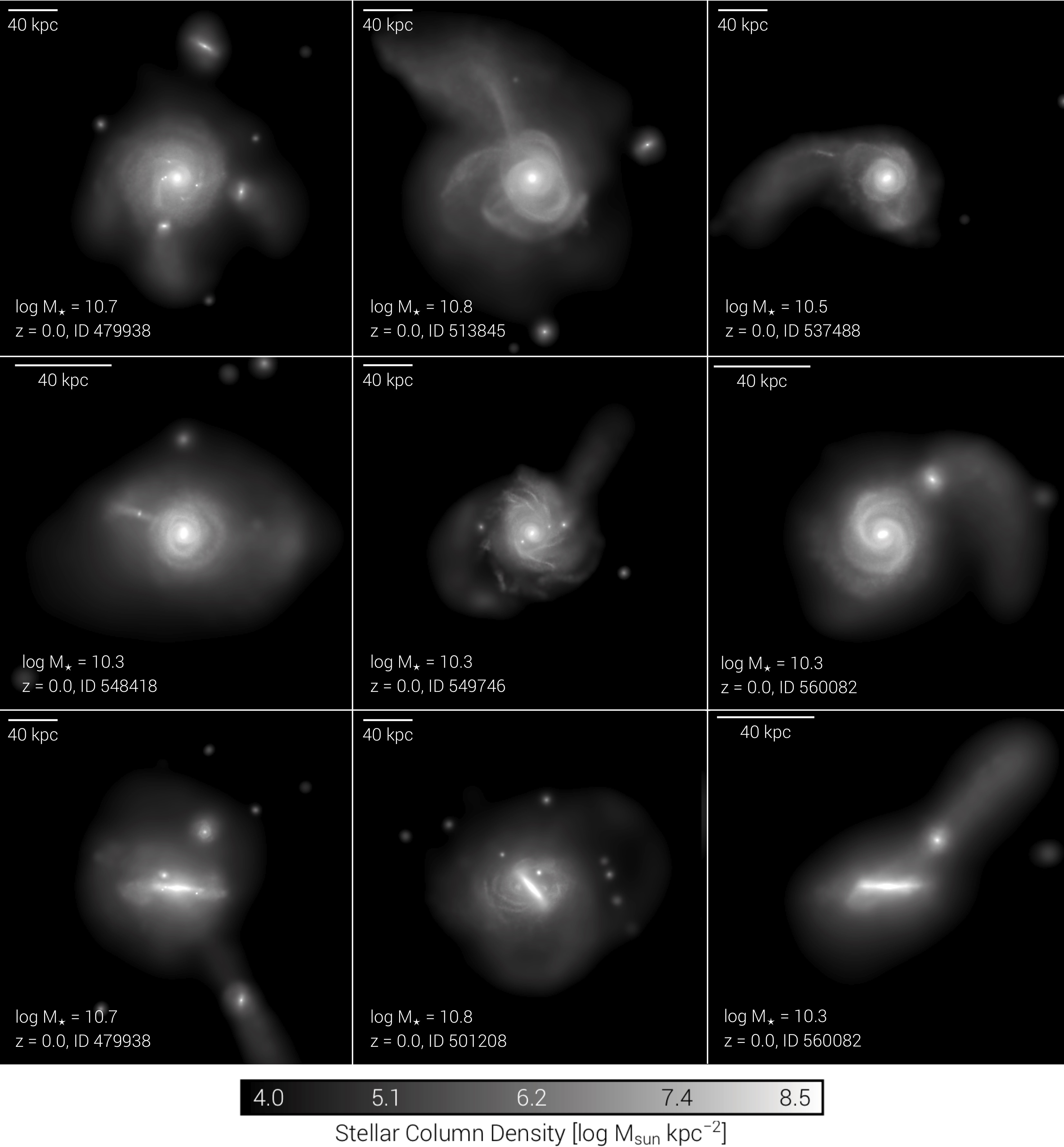}
\caption{Example galaxies from the TNG50 simulation. Images are in projected stellar column density above $10^4 M_{\odot}$ kpc$^{-2}$ and represent a subset of Milky Way-like galaxies featuring prominent stellar streams, either linear or curved. The stellar masses of the galaxies are in the range $\log_{10} M_{\star}/\mathrm{M_{\odot}} = 10.3 - 10.8$, as indicated. Notice that two of the galaxies are repeated in different projections. }
    
    \label{fig:TNG50_examples}
\end{figure*}
\subsubsection{The TNG50 hydrodynamical model}

It is reasonable to expect that other cosmological simulations, such as Illustris \citep{IllVogelsberger14a, IllGenel14, IllNelson2015} and IllustrisTNG \citep{TNGNelson18,naiman_2018,TNGPillepich18,TNGSpringel18,TNGMarinacci18,TNGDRNelson19}, may make different predictions for the properties of tidal stellar streams and the correlation of such properties with other galactic observables and intrinsic properties. Crucially, these differences arise not only from the different galaxy formation prescriptions used in various models, but also from fundamentally different choices in their underlying dynamical assumptions. 

For example, the \coco{} simulations assume pure $N$-body dynamics, with star formation treated semi-analytically\footnote{The gravitational dynamics of baryons 
\textit{are} included in the semi-analytic cooling, star formation and feedback models, albeit approximately.}. IllustrisTNG, on the other hand, includes not only gravity but also a self-consistent description of the gas hydrodynamics together with prescriptions for gas cooling and heating, formation and evolution of stellar particles, metal enrichment, galactic scale gas outflows due to star formation, and energy injections from super massive black holes \citep{TNGMethodsWeinberger17, TNGMethodsPillepich18}. Namely, the formation and evolution of massive thin disks of stars and gas are emerging phenomena in these models, and with them so are for example the star formation quenching of low-mass galaxies in dense environments and the assembly of galactic stellar haloes, whether of ex-situ or in-situ origin. The `higher order' processes that are ignored by semi-analytic models but which arise naturally in a hydrodynamical simulation are likely to be important in the detection and interpretation of low surface brightness structures. For example, satellites can interact with galactic disks, which may significantly affect the satellite disruption rate \citep{Errani2017, GarrisonKimmel2017}. Those interactions in turn may drive disk heating, warping and buckling, as well as asymmetric features, such as tidal arms. 

Large-volume galaxy simulations like Illustris and the TNG100 and TNG300 runs of the IllustrisTNG project\footnote{http://www.illustris-project.org/ and http://www.tng-project.org/} have already been queried to characterise the imprint of different accretion and merger histories on the stellar components of galaxies, from stellar disks to stellar haloes and intra-cluster light \citep[e.g.][]{Pillepich14, RodriguezGomez16}, including the formation and properties of shell galaxies \citep{Pop2017, Pop2018}. However, because of the somewhat limited numerical resolution, such studies have been limited to either global properties of stellar haloes or to the fine features in stellar haloes, such as streams and shells, of galaxies a few times more massive than the Milky Way.

Among the currently available cosmological hydrodynamical simulations of Milky Way analogues, the more recent TNG50 simulation \citep{TNG50Pillepich19, TNG50Nelson19} offers a unique combination of statistical sampling and numerical resolution.  TNG50 is the third and most ambitious realisation of the IllustrisTNG project. It simulates a 52 Mpc/side volume at numerical resolution similar to zoom-in simulations like Auriga (i.e. about 8$\times 10^4 \mathrm{M_{\odot}}$/stellar particle). The resulting volume contains $\sim200$ among Milky Way and Andromeda like galaxies at $z=0$, in addition to a few hundred more massive galaxies and thousands of dwarf galaxies, in both isolation or as satellites of more massive systems. The high mass resolution in this volume enables a detailed investigation of the present-day structural properties of galaxies, such as the thickness of  gaseous and stellar disks, perturbations in those disks due to interactions with satellites, and low surface brightness features in stellar halos. Fig.~\ref{fig:TNG50_examples} provides a snapshot of the diversity of stellar halo features of TNG50 simulated galaxies chosen among hundreds with stellar or halo mass similar to those of the Milky Way. In particular, represented is a subset of Milky Way-like galaxies (i.e. with disk-like stellar structures) featuring prominent stellar streams, either linear or curved. 

The comparison we anticipate here between the results of the {\it Stellar Stream Legacy Survey} and the outcome of the TNG50 simulation can offer great insight on both the galaxy formation model underlying the simulation as well as on the assembly histories of the observed galaxies.

\section{CONCLUSIONS}
\label{sec:conclusions}

In this paper we have introduced the {\it Stellar Stream Legacy Survey}, a new survey of low surface brightness features 
associated with the tidal disruption of dwarf galaxies around isolated massive galaxies in the local Universe. 
%
%
Our survey is designed to constrain the rate of occurrence of 
those features
and to provide robust distributions of their mass, morphology and colour. It is based on an optimised re-reduction of deep $g$, $r$ and $z$-band imaging data from the {\it DESI Legacy Imaging Surveys}, 
which are publicly available and cover about $20,000$ deg$^2$ (Sec.~\ref{sec:legacy}). Our parent sample of isolated galaxies comprises $\sim3100$ systems with $K$-band absolute magnitude M$_{K}< -19.6$, at distances of $30 < d < 100$~Mpc (Sec.~\ref{sec:surveydesign}).
The data we obtain  allow us to make direct, quantitative comparisons of tidal stream statistics to predictions from state-of-the-art $\Lambda$CDM cosmological simulations, and hence to constrain cosmological and sub-grid parameters controlling the assembly rates of massive galaxies and the stellar populations of dwarf satellites.

As a proof of concept of the survey, we have reported here the detection of 24 new stellar streams (Figs.~\ref{fig:Fig3A_July31} and \ref{fig:Fig3B_July31}). An initial list of these discoveries and previously known features
is presented in Table~\ref{tab:target_list}, which we will expand upon in future work. 
In this paper we have introduced the techniques that are used for the entire survey and have applied them to a subset of galaxies in our sample. We have demonstrated that:
\begin{itemize}

    \item Our custom reprocessing of data from the {\it DESI Legacy Imaging Surveys} has a $r$-band limiting depth of $\sim28.5 \pm 0.4\,\mathrm{mag\,arcsec^{-2}}$ ($3\sigma$ of the sky background variation measured in $10\times10\arcsec$ pixels). This is comparable to other ultra-deep observations of diffuse structures in the literature and sufficient for the detection of a significant sample of minor merger events associated with our target galaxies. In our proof of concept sample, these features are typically $1$ to $3\,\mathrm{mag\,arcsec^{-2}}$ brighter than our detection limit.
    
    \item The re-reduced Legacy Survey images provide sufficient signal-to-noise to obtain reliable colours for streams using our baseline method for aperture photometry. In our proof-of-concept sample, we find a range of colours $0.5\lesssim g-r \lesssim 0.8$, somewhat redder than the range of massive satellite galaxies in the SAGA survey. We extend and refine this analysis in subsequent work describing the full sample.
    
    \item Methods to distinguish features arising from dwarf galaxy accretion from those generated by major mergers and perturbations to thin galactic discs can be tested at the resolution of our survey (Sec.~\ref{sec:confusion}).
    
    \item Our approach offers a much larger sample size, is competitive with the {\it Euclid} Wide Survey and only marginally shallower than the 10-year stacked data expected from the\textit{Vera C. Rubin Telescope Legacy Survey of Space and Time}  (Sec.~\ref{sec:othersurveys}).
    
    \item The `particle spray' technique can rapidly generate a large library of bespoke $N$-body realisations of individual streams, with which to test both visual and automated classification methods. We use this technique in combination with convolutional neural networks to quantify stream morphologies in our sample for comparison with cosmological simulations. This approach can constrain 
    galactic accretion histories and satellite
    orbital distributions. We also use it to identify a large `gold standard' sample of thin streams, in order to constrain the potentials of their host galaxies and provide a new test of the galaxy--halo connection (Sec.~\ref{sec:discussion}).   
    
    \item The current generation of cosmological volume simulations is well matched to the spatial scale, resolution and sample size of our data. 
    Realistic mock images generated from these models enable quantitative comparisons with the results of our survey.
    (Sec.~\ref{sec:cosmosim}).   
    
\end{itemize}

The full dataset of the {\it Stellar Stream Legacy Survey} is presented in a forthcoming paper. Our analysis of its images acts as a pathfinder for future surveys with higher spatial resolution and greater depth, including those with {\it Euclid}, the \textit{Vera C. Rubin Observatory} and the {\it Nancy Grace Roman Space Telescope.}


\begin{acknowledgements}
We thank Eva K. Grebel, Facundo A. G\'omez, Aaron Romanowski and Emilio Galvez for useful comments. We thank Mireia Montes for her insight and expertise regarding the capabilities of Vera C. Rubin Telescope Survey. DMD acknowledges financial support from the Talentia Senior Program (through the incentive ASE-136) from Secretar\'\i a General de  Universidades, Investigaci\'{o}n y Tecnolog\'\i a, de la Junta de Andaluc\'\i a. DMD acknowledges funding from the State Agency for Research of the Spanish MCIU through the ``Center of Excellence Severo Ochoa" award to the Instituto de Astrof{\'i}sica de Andaluc{\'i}a (SEV-2017-0709) and AEI project (PDI2020-114581GB-C21/ AEI / 10.13039/501100011033). 
APC is supported by the Taiwan Ministry of Education (MoE) Yushan Fellowship and Taiwan National Science and Technology Council (NSTC) grant 109-2112-M-007-011-MY3. This work made use of HPC facilities at the National Tsing Hua University Center for Informatics and Computation in Astronomy (CICA), supported by grants from MoE and NSTC. This work has also been supported by the Spanish Ministry of Economy and Competitiveness (MINECO) under grant AYA2014-56795-P. This work was supported in part by World Premier International Research Center Initiative (WPI Initiative), MEXT, Japan. We acknowledge the usage of the HyperLeda database (http://leda.univ-lyon1.fr).
AP and DMD acknowledge support by the Deutsche Forschungsgemeinschaft (DFG, German Research Foundation) -- Project-ID 138713538 -- SFB 881 (``The Milky Way System'', subprojects A1 and A2). 
JR acknowledge financial support from the grants AYA2015-65973-C3-1-R and RTI2018-096228- BC31 (MINECO/FEDER, UE) and support from the State Research Agency (AEI-MCINN) of the Spanish Ministry of Science and Innovation under the grant "The structure and evolution of galaxies and their central regions" with
reference PID2019-105602GB-I00/10.13039/501100011033.
MA acknowledges support from the Spanish Ministry of Economy and Competi-tiveness (MINECO) under grant number AYA2016-76219-P and EU’s Horizon 2020 research and innovation programme under Marie Sklodowska-Curie grant agreement No 721463 to the SUNDIAL ITN.
W.J.P. has been supported by the Polish National Science Center project UMO-2020/37/B/ST9/00466.
CSF acknowledges support by the European Research Council (ERC) through
Advanced Investigator grant DMIDAS (GA 786910) and by the Science and
Technology Facilities Council (STFC) consolidated grant ST/P000541/1.
This work used the DiRAC@Durham facility managed by the Institute
for Computational Cosmology on behalf of the STFC DiRAC HPC Facility
(www.dirac.ac.uk). The equipment was funded by BEIS capital funding via
STFC capital grants ST/K00042X/1, ST/P002293/1, ST/R002371/1 and
ST/S002502/1, Durham University and STFC operations grant ST/R000832/1.
DiRAC is part of the National e-Infrastructure.

A.B. was supported by an appointment to the NASA Postdoctoral Program at the NASA Ames Research Center, administered by Universities Space Research Association under contract with NASA. 

This project uses data from observations at Cerro Tololo Inter-American Observatory, National Optical Astronomy Observatory, which is operated by the Association of Universities for Research in Astronomy (AURA) under a cooperative agreement with the National Science Foundation. We acknowledge support from the Spanish Ministry for Science, Innovation and Universities and FEDER funds through grant AYA2016-81065-C2-2. We also used data obtained with the Dark Energy Camera (DECam), which was constructed by the Dark Energy Survey (DES) collaboration. Funding for the DES Projects has been provided by
the U.S. Department of Energy, 
the U.S. National Science Foundation, 
the Ministry of Science and Education of Spain, 
the Science and Technology Facilities Council of the United Kingdom, 
the Higher Education Funding Council for England, 
the National Center for Supercomputing Applications at the University of Illinois at Urbana-Champaign, 
the Kavli Institute of Cosmological Physics at the University of Chicago, 
the Center for Cosmology and Astro-Particle Physics at the Ohio State University, 
the Mitchell Institute for Fundamental Physics and Astronomy at Texas A\&M University, 
Financiadora de Estudos e Projetos, Funda{\c c}{\~a}o Carlos Chagas Filho de Amparo {\`a} Pesquisa do Estado do Rio de Janeiro, 
Conselho Nacional de Desenvolvimento Cient{\'i}fico e Tecnol{\'o}gico and the Minist{\'e}rio da Ci{\^e}ncia, Tecnologia e Inovac{\~a}o, 
the Deutsche Forschungsgemeinschaft, 
and the Collaborating Institutions in the Dark Energy Survey. 
The Collaborating Institutions are 
Argonne National Laboratory, 
the University of California at Santa Cruz, 
the University of Cambridge, 
Centro de Investigaciones En{\'e}rgeticas, Medioambientales y Tecnol{\'o}gicas-Madrid, 
the University of Chicago, 
University College London, 
the DES-Brazil Consortium, 
the University of Edinburgh, 
the Eidgen{\"o}ssische Technische Hoch\-schule (ETH) Z{\"u}rich, 
Fermi National Accelerator Laboratory, 
the University of Illinois at Urbana-Champaign, 
the Institut de Ci{\`e}ncies de l'Espai (IEEC/CSIC), 
the Institut de F{\'i}sica d'Altes Energies, 
Lawrence Berkeley National Laboratory, 
the Ludwig-Maximilians Universit{\"a}t M{\"u}nchen and the associated Excellence Cluster Universe, 
the University of Michigan, 
{the} National Optical Astronomy Observatory, 
the University of Nottingham, 
the Ohio State University, 
the University of Pennsylvania, 
the University of Portsmouth, 
SLAC National Accelerator Laboratory, 
Stanford University, 
the University of Sussex, 
and Texas A\&M University.
Support for this work was provided by NASA through the NASA Hubble Fellowship grant \#HST-HF2-51466.001-A awarded by the Space Telescope Science Institute, which is operated by the Association of Universities for Research in Astronomy, Incorporated, under NASA contract NAS5-26555.
This work was partly done using GNU Astronomy Utilities (Gnuastro, ascl.net/1801.009) version \detectiongnuastrover. Work on Gnuastro has been funded by the Japanese MEXT scholarship and its Grant-in-Aid for Scientific Research (21244012, 24253003), the European Research Council (ERC) advanced grant 339659-MUSICOS, and from the Spanish Ministry of Economy and Competitiveness (MINECO) under grant number AYA2016-76219-P.

\end{acknowledgements}

%

%



\end{document}